\documentclass[12pt,
  amsmath,
  amssymb,
  superscriptaddress,
  longbibliography
]{revtex4-2}

\usepackage[T1]{fontenc}
\usepackage[utf8]{inputenc}
\usepackage{lmodern}
\usepackage{graphicx}
\usepackage{booktabs}
\usepackage{microtype}
\usepackage{xcolor}

\usepackage{hyperref}
\definecolor{linkblue}{RGB}{24,57,88}
\hypersetup{
  colorlinks=true,
  linkcolor=linkblue,
  citecolor=linkblue,
  urlcolor=linkblue,
  pdftitle={A Universal Axion in a Twin-QCD Sector},
  pdfsubject={Twin-QCD axion phenomenology, ultraviolet matching, and cosmology},
  pdfauthor={Stuart Raby, Scott Watson}
}

\graphicspath{{figures/}}

\newcommand{\gag}{g_{a\gamma\gamma}}
\newcommand{\mQ}{m_{a,\mathrm{QCD}}}
\newcommand{\Ra}{R_a}
\newcommand{\rL}{r_\Lambda}
\newcommand{\ralpha}{r_{\alpha}}
\newcommand{\Cg}{C_\gamma}
\newcommand{\GeV}{\mathrm{GeV}}
\newcommand{\eV}{\mathrm{eV}}
\newcommand{\meV}{\mathrm{meV}}
\newcommand{\mueV}{\mu\mathrm{eV}}
\newcommand{\flocal}{F_a^{\mathrm{local}}}
\newcommand{\fcos}{F_a^{\mathrm{cos}}}
\newcommand{\Mbar}{\overline M_{\mathrm{Pl}}}
\newcommand{\Neff}{N_{\mathrm{eff}}}
\newcommand{\Tosc}{T_{\mathrm{osc}}}

\newcommand{\dd}{\mathrm{d}}

\begin{document}

\title{A Universal Axion in a Twin-QCD Sector}


\author{Stuart Raby}
\affiliation{Department of Physics, Ohio State University, Columbus, Ohio, 43210, USA}

\author{Scott Watson}
\affiliation{Department of Physics, Syracuse University, Syracuse, NY 13244, USA}

\date{\today}

\begin{abstract}
Dark matter may consist of multiple components with distinct particle-physics
origins. We investigate a scenario in which a universal axion couples to
both visible and twin QCD, with twin baryons providing a possible dominant
dark-matter component. The presence of a second strongly interacting sector
can substantially enhance the axion mass while preserving its role in
resolving the strong-CP problem, provided the physical vacuum angles are
appropriately aligned. For a representative uniformly rescaled twin spectrum
with a confinement scale five times that of ordinary QCD, the axion mass is
enhanced by approximately a factor of 25.

Using lattice-informed QCD results and a two-temperature description of the
early universe, we investigate the cosmological evolution and relative
abundances of the dark-sector components. In a representative conditional numerical
benchmark, axions account for a small percentage of the observed dark
matter, while twin baryons could constitute the remainder for a suitable
primordial baryon asymmetry. We also examine the implications for additional
relativistic species, strong-CP alignment, and the interactions of stable
twin matter.

Our results provide a quantitative framework for exploring mixed
axion--twin-baryon dark matter and identify the particle-physics and
cosmological conditions necessary for a viable realization, including
constraints from the thermal history and formation of cosmic structure.
\end{abstract}

\maketitle
\newpage 

\section{Scope and assumptions}
\label{sec:scope}

Consider a twin sector with Standard Model gauge and matter content and a
universal axion that couples to both color groups. A motivation is the
visible/twin-sector structure of the Clemens--Raby F-theory construction
\cite{ClemensRaby2020}. The phenomenological benchmark examined here is
\begin{equation}
  \rL\equiv\frac{\Lambda_{\mathrm{QCD}}'}{\Lambda_{\mathrm{QCD}}}=5.
  \label{eq:rLambda}
\end{equation}
This note starts from the normalization of the proposed axion locus and
extends the analysis to the thermal history and composition of the dark
sector. The results are conditional effective-theory statements; they do not establish a complete construction or a fully tested cosmology.

The assumptions are that the axion is canonically normalized, the two color
anomaly coefficients are equal, the physical strong-CP minima are aligned,
and additional contributions to the axion potential are negligible. The
visible photon relation below also assumes $m_a\ll m_\pi$, large $f_a$
relative to hadronic scales, and no additional portal-induced correction
to the visible photon coupling. Shared-axion constructions illustrate both
the mass-enhancement mechanism and the importance of alignment
\cite{Berezhiani2001,Dunsky2023}.

The crucial ultraviolet requirement is stronger than equality of the
bare color anomalies. The canonically normalized light axion must
couple to both color groups with the specified relative coefficient,
and the two physical strong-CP angles, including quark mass determinant
phases and heavy thresholds, must coincide to neutron-EDM accuracy.
For the uniformly rescaled benchmark, $\chi'/\chi=625$, a small
mismatch produces
$\bar\theta_{\rm eff,vis}\simeq(625/626)
(\bar\theta-\bar\theta')$; the twin susceptibility does not suppress
the residual visible CP angle. Similarly, the conventional
$E/N=8/3$ is an explicit phenomenological input until the
light axion's electromagnetic anomaly is derived from the
specified construction. These questions are distinct from the
low-energy mass enhancement and from the cosmological population
calculations.

\subsection{Independent inputs and the meaning of a twin sector}
\label{sec:inputs}

The equality of the two gauge and matter representations should not be
confused with equality of their masses, temperatures, or baryon
asymmetries. In particular, a visible-sector copy is not necessarily an
exact exchange-symmetric vacuum. We use ``twin'' to denote the second
Standard-Model-like sector, without assuming the Higgs-sector relations
of a specific twin-Higgs model. The phenomenological inputs include
\begin{equation}
 \left\{f_a,\ \frac EN,\ \frac{\chi'}{\chi},\
 \frac{v'}v,\ \ralpha,\ \frac{T'}T,\ Y_{B'},\ \Theta_i\right\},
 \label{eq:input-set}
\end{equation}
where $v'/v$ controls masses only when the corresponding Yukawa couplings
are specified, $\ralpha\equiv\alpha'_{\mathrm{EM}}/\alpha_{\mathrm{EM}}$
is the low-energy twin-to-visible electromagnetic coupling ratio,
$Y_{B'}$ denotes a baryon-asymmetry yield, and $\Theta_i$
is the initial angular displacement from the common axion minimum.
Neither $T'/T$ nor $Y_{B'}$ follows from the axion anomaly coefficients.
Their precise conventions are defined below.

The F-theory construction of Ref.~\cite{ClemensRaby2020,Clemens:2019dts} motivates
studying visible and twin gauge sectors with independently adjustable
scales. Its explicit example involves a very different, high-scale
supersymmetric twin spectrum. Thus, that example is not itself a
derivation of the low-energy spectrum or of $\rL=5$ used here.
Obtaining this benchmark requires a separate choice of vacuum,
thresholds, and axion direction. This distinction is important when
turning a phenomenological relation into a compactification prediction.

There are three logically separate questions. The vacuum theory fixes
the mass and couplings; the cosmological history fixes the populations;
and the distributions of those populations fix the interpretation of
dark-matter searches. A successful treatment must answer all three
without replacing one by an assumption about another.

\section{Mass enhancement: susceptibility rather than confinement alone}
\label{sec:mass}

Normalize the color interactions as
\begin{equation}
  \mathcal L\supset\frac{a}{f_a}
  \left[
    \frac{\alpha_s}{8\pi}G_{\mu\nu}^{A}\widetilde G^{A\mu\nu}
    +\frac{\alpha_s'}{8\pi}G_{\mu\nu}^{\prime A}
          \widetilde G^{\prime A\mu\nu}
  \right].
  \label{eq:color-couplings}
\end{equation}
Equal anomaly coefficients in this convention do not require equal running
strong couplings. The zero-temperature mass follows from the curvature of
the two vacuum energies. For aligned minima,
\begin{equation}
  \begin{gathered}
    m_a^2f_a^2=\chi+\chi',\\[3pt]
    \boxed{\Ra\equiv\frac{m_a}{\mQ}
            =\sqrt{1+\frac{\chi'}{\chi}}.}
  \end{gathered}
  \label{eq:enhancement}
\end{equation}
Here $\chi$ and $\chi'$ are the visible and twin topological
susceptibilities, and $\mQ$ is the ordinary QCD axion mass at the
\emph{same} $f_a$ \cite{GrilliDiCortona2016,Dunsky2023}.

If all QCD dimensionful parameters are rescaled together, the dimensionless
ratios $m_q/\Lambda_{\mathrm{QCD}}$ are unchanged and dimensional analysis
gives $\chi'/\chi=\rL^4$. Thus,
\begin{equation}
  \Ra=\sqrt{1+5^4}=\sqrt{626}\simeq25.02.
  \label{eq:R25}
\end{equation}
The two contributions add in \emph{mass squared}; adding $1+5^2$ to obtain
26 is not the appropriate combination.

Identical gauge and matter content does not, however, enforce identical
mass ratios. To display this dependence, define
$\kappa\equiv m_u'/m_u=m_d'/m_d$. In the leading light-quark approximation,
\begin{equation}
  \chi\simeq\frac{m_um_d}{m_u+m_d}\,\Sigma,
  \qquad
  \Sigma=-\langle\bar q q\rangle,
  \label{eq:chiral}
\end{equation}
so the scaling estimate $\Sigma'/\Sigma\simeq\rL^3$ yields
\begin{equation}
  \frac{\chi'}{\chi}\simeq\kappa\rL^3,
  \qquad
  \Ra\simeq\sqrt{1+\kappa\rL^3}.
  \label{eq:chiral-scaling}
\end{equation}
The light-quark expression and its connection to mirror-sector scales
are discussed in Refs.~\cite{GrilliDiCortona2016,Berezhiani2001}.
For unchanged light-quark masses, $\kappa=1$ and $\rL=5$ give
$\Ra\simeq\sqrt{126}\simeq11.2$, rather than 25. These are scaling
benchmarks; quark thresholds and the applicable chiral regime must be
specified for a precision calculation. The equal-mass-ratio benchmark
also requires the relevant heavier-quark thresholds to rescale consistently.

In ongoing work on the underlying F-theory construction, the visible and
twin Yukawa couplings appear to coincide at the ultraviolet matching scale.
This is a preliminary property of the construction, not yet a completed
matching calculation. Equality of ultraviolet Yukawa couplings does not
imply equal low-energy fermion masses: schematically, in each sector
$m_f=y_f v/\sqrt{2}$ and $m_f'=y_f'v'/\sqrt{2}$, with the
renormalized Yukawas evaluated at the appropriate scales. Differences in
Higgs expectation values, renormalization-group running, and heavy-particle
thresholds can all affect $m_f'/m_f$. In particular, a choice
$v'/v\simeq5$ can motivate the uniform light-fermion rescaling only if
the running and matching preserve the requisite mass ratios; it does not
by itself produce $\Lambda'_{\mathrm{QCD}}/\Lambda_{\mathrm{QCD}}=5$.

\subsection{The effective potential and its range of validity}
\label{sec:potential}

Let $\bar\theta$ and $\bar\theta'$ be the physical angles after including
the phases of the quark mass determinants and any integrated-out colored
states. With the convention in Eq.~\eqref{eq:color-couplings}, define
\begin{equation}
 \Theta=\bar\theta+\frac{a}{f_a},\qquad
 \Delta=\bar\theta'-\bar\theta.
 \label{eq:angles}
\end{equation}
The vacuum potential has the general form
\begin{equation}
 V(a)=E(\Theta)+E'(\Theta+\Delta)+V_{\mathrm{br}}(a),
 \qquad
 \chi=\left.\frac{\partial^2 E}{\partial\Theta^2}\right|_0.
 \label{eq:general-potential}
\end{equation}
Here $V_{\mathrm{br}}$ denotes explicit breaking beyond the two color
anomalies. Equation~\eqref{eq:enhancement} follows by expanding around
$\Theta=\Delta=0$ and neglecting this extra term. It does not require
assuming that either QCD potential is an exact cosine.

For example, at leading order in the two-flavor chiral expansion,
the visible contribution, shifted to vanish at its minimum, is
\begin{equation}
 E(\Theta)-E(0)=m_\pi^2f_\pi^2
 \left[1-\sqrt{1-\frac{4z}{(1+z)^2}
                  \sin^2\!\left(\frac{\Theta}{2}\right)}\right],
 \qquad z=\frac{m_u}{m_d}.
 \label{eq:chiral-potential}
\end{equation}
Its curvature is $m_\pi^2f_\pi^2z/(1+z)^2$, while its large-angle shape
differs from a cosine. The twin contribution is obtained by replacing
its hadronic parameters, provided the twin theory remains in the
same chiral regime \cite{GrilliDiCortona2016}.
Large-angle misalignment therefore requires more information than
$\chi'(0)$ alone. In particular, a susceptibility fixes the quadratic
term but not all higher derivatives of the potential.

For nonuniform light-quark masses, the two-flavor scaling generalizes to
\begin{equation}
 \frac{\chi'}{\chi}\simeq\frac{\Sigma'}{\Sigma}
 \frac{m_u'm_d'/(m_u'+m_d')}{m_um_d/(m_u+m_d)}.
 \label{eq:unequal-light-quarks}
\end{equation}
Strange-quark effects, running-mass conventions, and higher-order
low-energy constants should be included in a quantitative evaluation.
If every twin quark is heavy compared with twin confinement, neither
Eq.~\eqref{eq:chiral} nor its simple $\kappa$ rescaling applies; the
appropriate susceptibility is that of the corresponding approximately
pure-glue theory, with its own matching coefficient. We are in the process of calculating the Yukawa couplings in our theory.   At the moment it is apparent that the Yukawa couplings in the visible and twin sectors are identical.

\subsection{Confinement-scale matching is a separate calculation}
\label{sec:thresholds}

The confinement scale is exponentially sensitive to the running gauge
coupling. At one loop, in a nonsupersymmetric interval with $n_f$ active
Dirac quarks,
\begin{equation}
 \alpha_s^{-1}(\mu)=\frac{b_{n_f}}{2\pi}
           \ln\!\frac{\mu}{\Lambda_{n_f}},
 \qquad b_{n_f}=11-\frac{2n_f}{3}.
 \label{eq:running}
\end{equation}
Continuity at a perturbative heavy-quark threshold gives
\begin{equation}
 \Lambda_{n_f-1}^{\,b_{n_f-1}}
 =m_h^{\,b_{n_f-1}-b_{n_f}}\Lambda_{n_f}^{\,b_{n_f}}.
 \label{eq:threshold-matching}
\end{equation}
Applying this relation at top, bottom, and charm yields, within the same
one-loop convention,
\begin{equation}
 \frac{\Lambda_3'}{\Lambda_3}
 =\left(\frac{\Lambda_6'}{\Lambda_6}\right)^{7/9}
  \left(\frac{m_t'm_b'm_c'}{m_tm_bm_c}\right)^{2/27}.
 \label{eq:lambda-thresholds}
\end{equation}
These equations make explicit the threshold sensitivity underlying
mirror-axion mass estimates \cite{Berezhiani2001}.

For identical six-flavor boundary couplings and a common rescaling
$m_h'=\kappa_hm_h$, Eq.~\eqref{eq:lambda-thresholds} would give
$\Lambda_3'/\Lambda_3=\kappa_h^{2/9}$, not $\kappa_h$.
Thus $\kappa_h=5$ produces only about $1.43$ in this restricted example.
The uniformly rescaled $\rL=5$ spectrum is perfectly meaningful as a
low-energy benchmark, but it is not obtained merely by multiplying all
quark masses by five while holding the ultraviolet coupling fixed.
Different boundary couplings or additional thresholds can change this
conclusion. Supersymmetric thresholds require the appropriate beta
functions and cannot be inserted into Eq.~\eqref{eq:lambda-thresholds}
without modification.

\section{The relation to the visible photon coupling}
\label{sec:photon}

At masses well below the pion scale, the visible photon coupling is
\begin{equation}
  \gag=\frac{\alpha_{\mathrm{EM}}}{2\pi f_a}\,\Cg,
  \qquad
  \Cg=\frac EN-1.92(4),
  \label{eq:photon-coupling}
\end{equation}
while the ordinary QCD axion mass is
\begin{equation}
  \mQ=5.70\,\mueV
  \left(\frac{10^{12}\,\GeV}{f_a}\right).
  \label{eq:qcd-mass}
\end{equation}
We use the central values from Ref.~\cite{GrilliDiCortona2016}. The ratio
$E/N$ in Eq.~\eqref{eq:photon-coupling} refers to the visible
electromagnetic and visible color anomalies in the normalization of
Eq.~\eqref{eq:color-couplings}; it is not obtained by summing the two
color-group anomalies in the denominator.

Using $m_a=\Ra\mQ$ to eliminate $f_a$ gives
\begin{equation}
  \boxed{
  |\gag|=2.04\times10^{-10}\,
  \frac{|\Cg|}{\Ra}
  \left(\frac{m_a}{\eV}\right)\GeV^{-1}.
  }
  \label{eq:master-line}
\end{equation}
This is the relation that should be placed on a mass--coupling plot.
For fixed $\Cg$, an enhancement $\Ra=25$ moves the ordinary QCD axion
locus rightward by a factor of 25 at fixed coupling, or downward by a
factor of 25 at fixed physical mass. The displacement in either coordinate
is $\log_{10}25\simeq1.40$ decades. In the supplied figure, successive
labeled mass prefixes, such as $\mueV$, $\meV$, and $\eV$, are separated
by \emph{three} decades.

\subsection{A conventional unified benchmark}
\label{sec:unified}

Complete $SU(5)$ multiplets with a common Peccei--Quinn charge within
each multiplet give the familiar benchmark $E/N=8/3$ with standard
hypercharge normalization \cite{GrilliDiCortona2016}. Its applicability
to the particular universal axion must be checked rather than assumed. Taking this value gives
$\Cg=8/3-1.92\simeq0.747$. With $\Ra=\sqrt{626}$, Eq.~\eqref{eq:master-line} becomes
\begin{equation}
  \boxed{
  |\gag|\simeq6.1\times10^{-12}
  \left(\frac{m_a}{\eV}\right)\GeV^{-1}.
  }
  \label{eq:benchmark-line}
\end{equation}
For example, at $f_a=10^{12}\,\GeV$,
\begin{equation}
  \begin{gathered}
    m_a\simeq1.43\times10^{-4}\,\eV,\\
    |\gag|\simeq8.7\times10^{-16}\,\GeV^{-1}.
  \end{gathered}
  \label{eq:benchmark-point}
\end{equation}
These numerical values are conditional predictions of the stated benchmark,
not a determination of the compactification parameters.

\subsection{Decay-constant benchmarks and other interactions}
\label{sec:coupling-benchmarks}

Table~\ref{tab:fa-benchmarks} gives a useful conversion between the
decay constant and the physical mass. All entries use the same intrinsic
visible coupling; no relic-density assumption is built into the table.
The two mass columns isolate the susceptibility dependence at fixed
$f_a$. Conversely, at a fixed measured mass, the larger-$\Ra$ benchmark
has a larger $f_a$ and consequently weaker interactions proportional
to $1/f_a$.

\begin{table*}[t]
 \centering
 \caption{Particle-physics benchmarks for $E/N=8/3$ and $\rL=5$.
 Masses and couplings use central values in
 Eqs.~\eqref{eq:photon-coupling}--\eqref{eq:qcd-mass}.
 The $\kappa=1$ column is a leading chiral scaling estimate.}
 \label{tab:fa-benchmarks}
 \begin{tabular*}{\textwidth}{@{\extracolsep{\fill}}cccc@{}}
 \toprule
 $f_a\ [\GeV]$ & $m_a\ [\eV]$, $\kappa=5$
 & $m_a\ [\eV]$, $\kappa=1$ & $|\gag|\ [\GeV^{-1}]$\\
 \midrule
 $10^9$  & $1.43\times10^{-1}$ & $6.40\times10^{-2}$ & $8.67\times10^{-13}$\\
 $10^{10}$ & $1.43\times10^{-2}$ & $6.40\times10^{-3}$ & $8.67\times10^{-14}$\\
 $10^{11}$ & $1.43\times10^{-3}$ & $6.40\times10^{-4}$ & $8.67\times10^{-15}$\\
 $10^{12}$ & $1.43\times10^{-4}$ & $6.40\times10^{-5}$ & $8.67\times10^{-16}$\\
 $10^{13}$ & $1.43\times10^{-5}$ & $6.40\times10^{-6}$ & $8.67\times10^{-17}$\\
 \bottomrule
 \end{tabular*}
\end{table*}

For an unbroken twin electromagnetic group, write
\begin{equation}
 g_{a\gamma'\gamma'}=
   \frac{\alpha'_{\mathrm{EM}}}{2\pi f_a}\,C_{\gamma'}.
 \label{eq:twin-photon-coupling}
\end{equation}
Equal color anomalies do not imply $C_{\gamma'}=\Cg$ or
$\alpha'_{\mathrm{EM}}=\alpha_{\mathrm{EM}}$. The twin electromagnetic
anomaly, its quark-mass ratios, and possible kinetic mixing must be
matched separately. The parameter $\ralpha$, defined in
Eq.~\eqref{eq:input-set}, depends on the ultraviolet gauge couplings,
charged-particle thresholds, and their running. Distinct unified gauge
couplings can arise from independent geometric moduli in the motivating
F-theory construction~\cite{ClemensRaby2020}; a larger twin GUT scale
alone does not fix the sign or magnitude of $\ralpha-1$. Equality is an additional
benchmark when the relevant charge assignments and dimensionless
hadronic parameters coincide. Twin-photon interactions can affect the
thermal history even though the displayed experimental plane contains
only the visible coupling \cite{Dunsky2023}.

Derivative fermion interactions may similarly be written as
\begin{equation}
 \mathcal L_{a\psi}=
 \frac{\partial_\mu a}{2f_a}\sum_\psi C_\psi\,
       \bar\psi\gamma^\mu\gamma_5\psi.
 \label{eq:fermion-couplings}
\end{equation}
Their coefficients are not fixed by $E/N$.
A model below an axion--photon exclusion can still be tested through
other interactions. Conversely, suppressing the visible photon anomaly
does not remove the color coupling used to generate the mass.

\section{The intrinsic axion--photon locus}
\label{sec:figure}

The visible mass--coupling relation in Eq.~\eqref{eq:master-line}
refers to the intrinsic particle interaction, not the abundance-weighted
sensitivity of an individual dark-matter experiment. For the unified
benchmark $E/N=8/3$ and uniform rescaling $\rL=5$, it predicts
Eq.~\eqref{eq:benchmark-line}. At $m_a=1\,\meV$, the ordinary QCD
axion has $|\gag|\simeq1.5\times10^{-13}\,\GeV^{-1}$ in this
benchmark, while the uniformly rescaled twin sector gives
$6.1\times10^{-15}\,\GeV^{-1}$. The alternative leading chiral
estimate $\Ra\simeq11.2$ gives approximately
$1.4\times10^{-14}\,\GeV^{-1}$ at the same mass.

\begin{figure*}[t]
 \centering
 \includegraphics[width=0.97\textwidth]{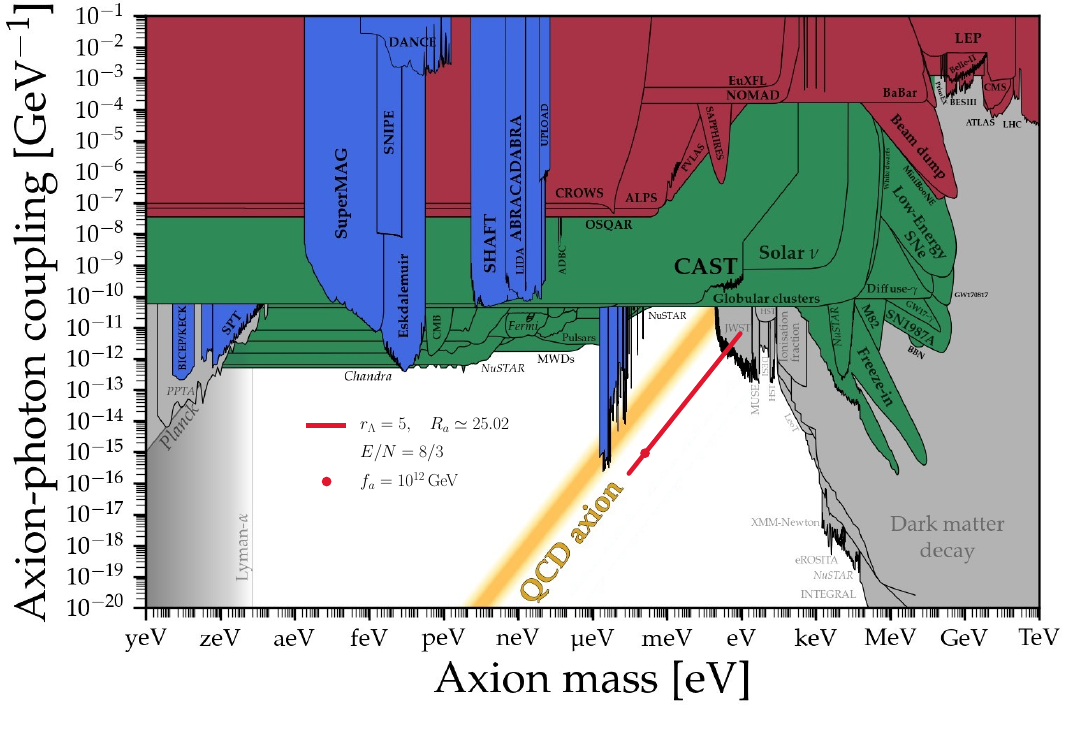}
\caption{Experimental and astrophysical constraints on the
axion--photon coupling, adapted from the FIPS Workshop 2022
compilation by C.~A.~J.~O'Hare, available through the
AxionLimits project~\cite{OHareAxionLimits}.
The superimposed red line represents our universal-axion
benchmark with $r_\Lambda=5$, corresponding to an axion
mass enhancement $R_a\simeq25$.
The red point indicates the benchmark with
$f_a=10^{12}\,\mathrm{GeV}$.
Individual exclusion regions depend on the assumptions
of the corresponding experiments, including the assumed
local axion dark-matter abundance where applicable.}
\label{fig:supplied}
\end{figure*}

Figure~\ref{fig:supplied} displays an illustrative compilation of
axion--photon constraints attributed to the AxionLimits resource of
O'Hare~\cite{OHareAxionLimits}, with the uniform-scaling benchmark
highlighted in red. Its marked point at $f_a=10^{12}\,\GeV$
agrees, to the resolution of the plot, with
Eq.~\eqref{eq:benchmark-point}. The slope and normalization
of the red segment are therefore consistent with
Eq.~\eqref{eq:benchmark-line}; an additional near-cancellation in
$\Cg$ is not implied by the plotted benchmark. The background compilation draws on limits with different statistical,
cosmological, and astrophysical assumptions. Its exact plot version has not
been established, so the overlay should not be used to assign precision
exclusion statements to individual regions~\cite{OHareAxionLimits}.

\begin{figure}[t]
 \centering
 \includegraphics[width=0.94\columnwidth]{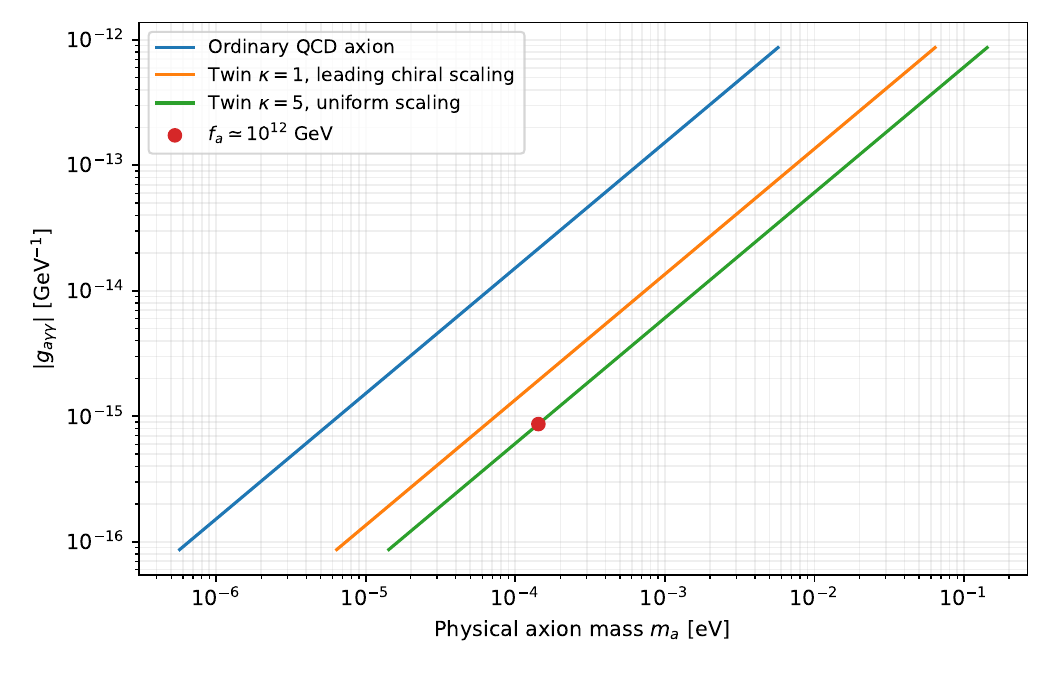}
 \caption{Intrinsic axion--photon loci for $E/N=8/3$: ordinary QCD,
 the $\rL=5$ leading chiral benchmark with $\kappa=1$, and the
 uniformly rescaled $\kappa=5$ benchmark. The marked point
 corresponds to $f_a=10^{12}\,\GeV$ in the uniform-scaling case.
 These curves show intrinsic couplings and not the relic abundance;
 the $\kappa=1$ relation is approximate, and neither case is
 derived from a specified compactification.}
 \label{fig:theory-loci}
\end{figure}

For comparison, Fig.~\ref{fig:theory-loci} shows only the
particle-physics relations, without overlaid experimental regions.
At fixed $f_a$, the enhanced susceptibility increases the physical
axion mass while leaving the assumed visible photon coupling unchanged.
At fixed $m_a$, the coupling is reduced by $1/\Ra$ relative to the
ordinary QCD axion when $\Cg$ is held fixed.

\section{Intrinsic coupling versus dark-matter abundance}
\label{sec:abundance}

The vertical axis of Fig.~\ref{fig:supplied} is the intrinsic particle
coupling, not an abundance-weighted signal strength. This distinction is
important if twin baryons supply most of the dark matter. Coupling
compilations combine laboratory, stellar, cosmological, and dark-matter
searches whose assumptions are not interchangeable
\cite{OHareAxionLimits}.

For fixed apparatus and halo velocity distribution, the haloscope signal
power scales schematically as
\begin{equation}
  P\propto |\gag|^2\frac{\rho_a}{m_a}.
  \label{eq:power}
\end{equation}

Defining
\begin{equation}
  \flocal\equiv\frac{\rho_a}{\rho_{\mathrm{DM}}},
  \label{eq:fraction}
\end{equation}
the rescaling of an all-axion-dark-matter coupling limit is
\begin{equation}
  g_{\mathrm{lim}}(\flocal)
  =\frac{g_{\mathrm{lim}}(1)}{\sqrt{\flocal}}.
  \label{eq:limit-rescaling}
\end{equation}
This relation holds with other experimental and halo assumptions fixed.
It changes the inferred exclusion, not the intrinsic model line.

For example, under the illustrative assumption that only three percent
of the local halo density is in axions, $\flocal=0.03$,
Eq.~\eqref{eq:limit-rescaling} weakens an all-axion haloscope coupling
limit by a factor $1/\sqrt{0.03}\simeq5.8$. This assumes a specified
halo density and does not change the intrinsic line in
Fig.~\ref{fig:theory-loci}. The local axion fraction need not equal
the cosmological fraction if the dominant twin matter is dissipative.

Stellar-production constraints do not acquire the same halo-abundance
rescaling. Cosmological and decay limits require a separate calculation
of the axion population, lifetime, and decay channels. In particular,
a common axion can couple to twin photons as well as visible photons;
mirror-axion cosmologies illustrate the consequences for radiation and
entropy transfer \cite{Dunsky2023}. A white region in the supplied
compilation should therefore not be read as proof that a complete
twin-sector cosmology is viable.

\section{Ultraviolet matching, strong-CP alignment, and axion quality}
\label{sec:matching}

The first ultraviolet matching task is to determine the coefficients of
$aG\widetilde G$, $aG'\widetilde G'$, and $aF\widetilde F$ for the canonically normalized light axion. Together with the twin quark
masses and thresholds, these determine the quantities that fix the locus:
$f_a$, $E/N$, and $\chi'/\chi$. The twin-photon coupling and any additional
axion potential must also be checked. An equal color coupling by itself
does not specify the visible photon coefficient.

\subsection{Alignment of the physical strong-CP angles}

Strong-CP alignment remains a separate consistency condition. Near the
aligned limit, expanding the two vacuum energies gives
\begin{equation}
  \bar\theta_{\mathrm{eff,vis}}
  =\frac{\chi'}{\chi+\chi'}
   \bigl(\bar\theta-\bar\theta'\bigr).
  \label{eq:theta-mismatch}
\end{equation}
Thus, when $\chi'$ dominates, a mismatch of the physical strong-CP angles
is not diluted. Their alignment, including mass-matrix phases, must
follow from the underlying theory rather than from the mass formula
alone \cite{Berezhiani2001,Dunsky2023}.

This result can be seen directly by expanding
Eq.~\eqref{eq:general-potential}:
\begin{equation}
 V=\frac12\chi\Theta^2+\frac12\chi'(\Theta+\Delta)^2+
   \text{higher-order terms}.
 \label{eq:quadratic-alignment}
\end{equation}
Minimization gives
$\Theta_{\min}=-\chi'\Delta/(\chi+\chi')$, reproducing
Eq.~\eqref{eq:theta-mismatch}. For $\chi'/\chi=625$, the coefficient
$625/626$ is essentially unity. A representative strong-CP target
$|\Theta_{\min}|\lesssim10^{-10}$ therefore requires comparably accurate
alignment of the physical angles. This order-of-magnitude target reflects
the neutron electric-dipole constraint and its hadronic interpretation;
it is not a new precision extraction of that constraint
\cite{GrilliDiCortona2016,Abel2020}.

Equal bare theta parameters are insufficient if the quark determinants
have different phases. The matching must include
\begin{equation}
 \bar\theta=\theta_{\mathrm{YM}}+\arg\det(M_uM_d)
          +\delta\theta_{\mathrm{heavy}},
 \label{eq:theta-physical}
\end{equation}
with a corresponding primed expression. Exchange or CP symmetries may
control these terms, but symmetry breaking and complex heavy thresholds
must be assessed in the specified model. There is one dynamical axion
variable and, generically, two independent CP angles; simultaneous
relaxation is not automatic.

\subsection{Additional sources of the axion potential}
\label{sec:quality}

Choose the angular origin at the aligned QCD minimum and parameterize
one explicit-breaking contribution by
\begin{equation}
 V_{\mathrm{br}}=-\Lambda_{\mathrm{br}}^4\cos(k\Theta+\delta).
 \label{eq:quality-potential}
\end{equation}
The allowed values of $k$ depend on the actual axion periodicity.
For a perturbatively small contribution and small $\Delta$,
\begin{equation}
 \Theta_{\min}\simeq
 -\frac{\chi'\Delta+k\Lambda_{\mathrm{br}}^4\sin\delta}
        {\chi+\chi'}.
 \label{eq:quality-shift}
\end{equation}
Thus the condition is on the \emph{slope} of the unwanted potential,
not only its mass contribution. In the absence of a tuned cancellation,
\begin{equation}
 |k|\Lambda_{\mathrm{br}}^4|\sin\delta|
 \lesssim \theta_{\max}(\chi+\chi'),
 \qquad \theta_{\max}\sim10^{-10}.
 \label{eq:quality-bound}
\end{equation}
A vanishing slope does not by itself guarantee an irrelevant correction
to the mass: $k^2\Lambda_{\mathrm{br}}^4\cos\delta$ must also be compared
with $\chi+\chi'$.

Using $\sqrt\chi=\mQ f_a=5.70\times10^{-3}\GeV^2$ gives
$\chi+\chi'\simeq2.03\times10^{-2}\GeV^4$ for the uniformly rescaled
benchmark. The right side of Eq.~\eqref{eq:quality-bound} is then about
$2.0\times10^{-12}\GeV^4$. Enhancement strengthens the restoring
potential by a factor of 626 relative to visible QCD alone, but does
not justify ignoring arbitrary ultraviolet breaking. The improvement
of axion quality from an aligned mirror potential is a separate issue
from the alignment itself \cite{Hook2020}.

\subsection{Canonical axion normalization and anomaly matching}
\label{sec:canonical-matching}

A geometric axion coordinate need not be the canonically normalized light
mass eigenstate. Let dimensionless periodic coordinates $\vartheta^I$
have kinetic matrix $K_{IJ}$, with mass dimension two, and interactions
\begin{equation}
 \mathcal L\supset\frac12K_{IJ}\partial_\mu\vartheta^I
                            \partial^\mu\vartheta^J
 +\sum_X\frac{\alpha_X}{8\pi}k_{XI}\vartheta^I
                         X_{\mu\nu}\widetilde X^{\mu\nu}.
 \label{eq:axion-kinetic-matrix}
\end{equation}
If all orthogonal modes can be integrated out, parameterize the light
direction by $\vartheta^I=v^Ia+\text{constant}$, with
\begin{equation}
 v^IK_{IJ}v^J=1,\qquad c_X=\sum_I k_{XI}v^I.
 \label{eq:canonical-projection}
\end{equation}
The coefficients $c_X$ have inverse-mass dimension. Choosing $c_3>0$
sets $f_a=1/c_3$, and the benchmark requires $c_{3'}=c_3$.
The visible ultraviolet electromagnetic anomaly ratio is
$c_{\mathrm{EM}}/c_3$ in the stated normalization. A sum over two
independent color groups is not the electromagnetic anomaly denominator.

More generally, around a common minimum the color contribution satisfies
\begin{equation}
 m_a^2 f_a^2=\chi+
       \left(\frac{c_{3'}}{c_3}\right)^2\chi'.
 \label{eq:unequal-anomalies}
\end{equation}
This is useful for testing whether small changes in the light eigenvector
alter the benchmark. If another axion is comparably light, the
single-field reduction is insufficient: the full mass matrix must be
diagonalized, and the photon coupling projected onto each eigenstate.
Likewise, stabilization must not give the putative QCD axion a larger,
misaligned potential. These are matching questions, not quantities
that can be fixed by adjusting the line on a coupling plot.

\section{Two-temperature thermal history and extra radiation}
\label{sec:thermal}

\subsection{Temperature ratio, entropy, and the expansion rate}

Let $T$ and $T'$ denote the visible and twin photon/plasma temperatures
at the same cosmic time, and define $\xi(T)=T'/T$.
Use $g_*$ for energy and $g_{*s}$ for entropy degrees of freedom, with
primed functions for the twin sector. After species decouple, these
functions include their individual temperatures relative to the sector's
photon temperature. During radiation domination, neglecting the axion
and nonrelativistic components in the background,
\begin{equation}
 H(T)=\sqrt{\frac{\pi^2}{90}}
 \frac{T^2}{\Mbar}
 \left[g_*(T)+g_*'(T')\xi^4(T)\right]^{1/2},
 \qquad \Mbar\simeq2.435\times10^{18}\GeV.
 \label{eq:hubble-two-temperature}
\end{equation}
The twin-sector clock is therefore not given by a universe filled only
with twin radiation at $T'$. For a colder, subdominant twin bath, the
visible bath largely determines $H$.

In the absence of intersector transfer and late entropy production,
$sA^3$ and $s'A^3$ are separately conserved, where $A(t)$ is the
cosmological scale factor, distinct from the axion field $a$.
With $s=(2\pi^2/45)g_{*s}T^3$, this gives
\begin{equation}
 \xi(T)=\xi_R
 \left[\frac{g_{*s}(T)}{g_{*s}(T_R)}\right]^{1/3}
 \left[\frac{g_{*s}'(T_R')}{g_{*s}'(T')}\right]^{1/3}.
 \label{eq:xi-entropy}
\end{equation}
The reheating values carry a subscript $R$. Equation~\eqref{eq:xi-entropy}
is implicit when $g_{*s}'$ depends on $T'=\xi T$.
It shows why the temperature ratio at axion oscillation cannot in general
be identified with the late photon temperature ratio. In the uniformly
rescaled QCD limit the two confinement crossovers occur at different
visible temperatures, which can temporarily change $\xi$.

For instantaneous reheating by a common parent with branching fractions
$B$ and $B'$ into internally equilibrated radiation baths,
\begin{equation}
 \xi_R^4=\frac{B'}{B}\frac{g_*(T_R)}{g_*'(T_R')}.
 \label{eq:reheating-ratio}
\end{equation}
Here $B$ and $B'$ are, respectively, the fractions of the parent energy
deposited into visible and twin radiation, so
$\rho_R'/\rho_R=B'/B$. Since
$\rho_R=(\pi^2/30)g_*(T_R)T_R^4$ and similarly for the twin bath,
the factor in Eq.~\eqref{eq:reheating-ratio} is $B'/B$, not its inverse.
Thus a small $\xi_R$ can encode asymmetric reheating. The subsequent
entropy history still has to be evolved; Eq.~\eqref{eq:reheating-ratio}
is not generally a relation between branching fractions and today's
photon temperatures.

\subsection{The twin contribution to the radiation budget}
\label{sec:dark-radiation}

After visible electron--positron annihilation, define extra radiation by
\begin{equation}
 \rho_{\mathrm{rad}}=\rho_\gamma
 \left[1+\frac78\left(\frac4{11}\right)^{4/3}\Neff\right].
 \label{eq:neff-definition}
\end{equation}
Suppose the relativistic twin population consists of a massless twin
photon and $N_\nu'$ neutrino species, each with two populated helicity
states. Put $r_\nu'=T_\nu'/T_\gamma'$ and
$\xi_\gamma=T_\gamma'/T_\gamma$. Their contribution follows directly
from their thermal energy densities:
\begin{equation}
 \Delta\Neff=
 \left[\frac87\left(\frac{11}{4}\right)^{4/3}
 +N_\nu'\left(\frac{r_\nu'}{(4/11)^{1/3}}\right)^4\right]
 \xi_\gamma^4.
 \label{eq:delta-neff}
\end{equation}
For three effectively massless twin neutrinos with
$r_\nu'=(4/11)^{1/3}$, the coefficient is approximately $7.40$.
This is the late-time photon-temperature convention; it should not be
confused with the familiar pre-annihilation coefficient near $6.14$.
The neutrino temperature, number of populated helicities, and whether
the neutrinos remain relativistic must all be checked in the actual
thermal history.

As an illustration, $\xi_\gamma=0.40$ gives
$\Delta\Neff\simeq0.19$ in the restricted radiation model above.
This is an outcome of the stipulated particle content and temperatures,
not a fitted value for the twin sector. Planck supplies a useful
reference for radiation constraints \cite{Planck2018Parameters}.
For a specified extension with extra free-streaming radiation, ACT DR6
combined with other cosmological data reports
$N_{\rm eff}=2.86\pm0.13$ \cite{Calabrese2025ACT}. Consequently
$\Delta\Neff\simeq0.19$ is not automatically allowed, nor is it
necessarily excluded by that number alone. The relevant likelihood
must take account of which twin species remain relativistic and
free-streaming, their temperatures, and any interactions between twin
radiation and twin baryons \cite{Bansal2022,Trendafilova2026Interacting}.
A viable radiation budget is necessary but not sufficient for a
viable twin-baryon cosmology.

\subsection{Can the universal axion reconnect the two baths?}
\label{sec:portal}

Separate entropy conservation must be tested against the same
interactions that define the axion. Schematically, energy transfer
between the baths obeys
\begin{align}
 \dot\rho+3H(\rho+p)&=-\mathcal Q,\\
 \dot\rho'+3H(\rho'+p')&=+\mathcal Q,
 \label{eq:energy-transfer}
\end{align}
where $\mathcal Q$ is positive for net transfer to the twin bath.
A non-negligible axion population requires its own kinetic equation and
consistent source terms rather than counting its energy in both baths.
The shared-axion portal can materially affect the entropy budget
\cite{Dunsky2023}.

At temperatures well above confinement and below the effective-theory
cutoff, dimensional estimates distinguish two possibilities. Thermal
axion production in one bath scales parametrically as
\begin{equation}
 \Gamma_{a,\mathrm{vis}}\sim
 c_s\frac{\alpha_s^3T^3}{f_a^2},
 \label{eq:thermal-axion-rate}
\end{equation}
with a corresponding twin rate. Off-shell intersector scattering
through two anomalous vertices instead has the schematic scaling
\begin{equation}
 \Gamma_{gg\to g'g'}\sim
 c_p\frac{\alpha_s^2\alpha_s'^2T^5}{f_a^4}
 \quad (T\gg m_a).
 \label{eq:off-shell-rate}
\end{equation}
These are power-counting estimates, not thermally averaged collision
integrals; $c_s$ and $c_p$ contain phase space, numerical anomaly factors,
and plasma effects. Sequential production and absorption of real axions
can dominate over off-shell scattering. Consequently, checking only
Eq.~\eqref{eq:off-shell-rate} is not sufficient to establish thermal
isolation. The appropriate comparison is between energy-transfer
rates and $H$ throughout the specified reheating history.

At low temperatures, visible and twin photons can also contribute to
the axion lifetime. For the convention
$\mathcal L\supset- g_{a\gamma\gamma}aF\widetilde F/4$, the two-body width is
\begin{equation}
 \Gamma_{a\to\gamma\gamma}=
 \frac{|\gag|^2m_a^3}{64\pi},\qquad
 \Gamma_{a\to\gamma'\gamma'}=
 \frac{|g_{a\gamma'\gamma'}|^2m_a^3}{64\pi}
 \label{eq:decay-widths}
\end{equation}
for massless final photons. Additional kinematically allowed channels
must be included before assigning a lifetime or decay branching ratio.

\section{Axion production in the two-temperature universe}
\label{sec:relic}

\subsection{Finite-temperature susceptibility and the onset of motion}

For aligned minima and negligible explicit breaking, the instantaneous
small-amplitude mass is
\begin{equation}
 m_a^2(T,T')=\frac{\chi(T)+\chi'(T')}{f_a^2}.
 \label{eq:finite-temperature-mass}
\end{equation}
Under the \emph{full} uniform rescaling used to obtain
Eq.~\eqref{eq:R25}, the finite-temperature relation is
\begin{equation}
 \chi'(T')=\rL^4\chi(T'/\rL).
 \label{eq:finite-temperature-scaling}
\end{equation}
This requires the dimensionless QCD parameters and all relevant
threshold ratios to match. It is not valid merely because
$\Lambda'_{\mathrm{QCD}}=5\Lambda_{\mathrm{QCD}}$.
Lattice calculations of $\chi(T)$ and the equation of state provide the
visible-sector input \cite{Borsanyi2016}; the changed-mass benchmark
requires its own twin susceptibility or a controlled approximation.

The homogeneous field obeys
\begin{equation}
 \ddot\Theta+3H\dot\Theta+
       \frac{1}{f_a^2}\frac{\partial V(\Theta;T,T')}{\partial\Theta}=0.
 \label{eq:misalignment-eom}
\end{equation}
In the small-angle regime, the last term is $m_a^2(T,T')\Theta$.
The conventional onset estimate is
\begin{equation}
 m_a\bigl(\Tosc,\xi(\Tosc)\Tosc\bigr)\simeq3H(\Tosc).
 \label{eq:oscillation-onset}
\end{equation}
This is an operational estimate of a continuous transition, not an
exact boundary. The numerical solution fixes the corresponding
order-unity onset correction.

A colder twin sector can turn on its potential at a larger visible
temperature and start axion oscillations earlier. This mechanism is
already present in mirror-axion cosmology \cite{Giannotti2005}.
However, $\Ra$ describes a \emph{zero-temperature} enhancement, while
Eq.~\eqref{eq:oscillation-onset} depends on both temperature-dependent
susceptibilities. Replacing the ordinary mass everywhere by
$\Ra\mQ$ generally misses this distinction.

\subsection{Comoving number and the relic abundance}
\label{sec:yield}

Once $m_a\gg H$ and $|\dot m_a|\ll m_a^2$, the action variable of the
nearly harmonic oscillator is adiabatic. In this regime,
\begin{equation}
 n_a\simeq\frac{\langle\rho_a\rangle}{m_a},\qquad
 A^3n_a\simeq\mathrm{constant},\qquad
 Y_a\equiv\frac{n_a}{s}\simeq\mathrm{constant}
 \label{eq:adiabatic-yield}
\end{equation}
when visible comoving entropy is conserved. The onset estimate gives
\begin{equation}
 Y_a\simeq\mathcal C_{\mathrm{osc}}
 \frac{f_a^2\Theta_i^2\,
       m_a(\Tosc,T'_{\mathrm{osc}})}{2s(\Tosc)}
 \mathcal F_{\mathrm{anh}}(\Theta_i).
 \label{eq:yield-estimate}
\end{equation}
Here $\mathcal C_{\mathrm{osc}}$ accounts for matching through the onset
and $\mathcal F_{\mathrm{anh}}\to1$ at small angle. Their separation is
only a useful parameterization, not two independent physical sources.
A large-angle evolution should instead use the specified full potential.

For a stable, nonrelativistic population today,
\begin{equation}
 \Omega_a h^2=\frac{s_0}{\rho_c/h^2}\,m_a(0)Y_a,
 \qquad
 \fcos=\frac{\Omega_a}{\Omega_{\mathrm{DM}}}.
 \label{eq:relic-density}
\end{equation}
The factor $m_a(0)$ in this expression is important. An earlier onset
reduces the comoving number, but a larger final mass increases the energy
per particle. Those two effects need not yield a net suppression.
If visible comoving entropy increases by a factor
$D=S_{\mathrm{vis,after}}/S_{\mathrm{vis,before}}$ after number conservation
is established, $Y_a$ is divided by $D$. Oscillations during the
decaying matter-dominated era itself require solving its background;
postulating a dilution factor alone is not sufficient.

Thermally produced axions, decays of heavy fields, and string--wall
production are not included in Eq.~\eqref{eq:yield-estimate}.
They must be added with their momentum distributions when present.
A relativistic component contributes to radiation rather than to the
cold fraction used in a haloscope interpretation.

\subsection{A controlled high-temperature scaling limit}
\label{sec:analytic-relic}

A useful analytic limit exposes the competition just described.
Approximate the ordinary high-temperature mass by
\begin{equation}
 m_{a,\mathrm{QCD}}(T)=m_{Q,0}(T_c/T)^n,
 \qquad n>0,
 \label{eq:power-law-mass}
\end{equation}
where $m_{Q,0}=\mQ(T=0)$ and $T_c$ is a matching scale.
Thus the corresponding susceptibility scales as $T^{-2n}$.
Assume constant $\xi$, constant visible $g_*$ and $g_{*s}$, negligible
twin contribution to $H$, small initial angle, and oscillation onset
above the respective crossover in both the ordinary and twin theories.
If the twin term dominates at onset, Eq.~\eqref{eq:finite-temperature-scaling}
implies
\begin{equation}
 m_a(T,\xi T)\simeq
 m_{Q,0}\rL^{\,2+n}\xi^{-n}(T_c/T)^n.
 \label{eq:power-law-twin-mass}
\end{equation}
Let $T_{\mathrm{osc,Q}}$ denote the ordinary-QCD onset at the \emph{same}
$f_a$. Dividing the two onset equations gives
\begin{equation}
 \frac{\Tosc}{T_{\mathrm{osc,Q}}}
 \simeq \rL\,\xi^{-n/(n+2)}.
 \label{eq:onset-ratio}
\end{equation}
Since $m_a(\Tosc)=3H\propto\Tosc^2$ and $s\propto\Tosc^3$,
Eq.~\eqref{eq:yield-estimate} yields $Y_a\propto f_a^2/\Tosc$.
At equal initial angle and equal onset corrections, the relic ratio is
therefore
\begin{equation}
 \boxed{\frac{\Omega_a}{\Omega_{a,\mathrm{QCD}}}
 \simeq\frac{\Ra}{\rL}\,\xi^{n/(n+2)}
 \simeq\rL\,\xi^{n/(n+2)}\quad(\rL^4\gg1).}
 \label{eq:relic-ratio}
\end{equation}
These are analytical estimates derived under the assumptions above,
not the output of a numerical parameter scan.

For $n=4$, $\rL=5$, and $\xi=0.40$, Eq.~\eqref{eq:relic-ratio}
is about $2.7$: the axion abundance is \emph{larger} than in the
ordinary-QCD comparison despite the colder twin bath. Net suppression
in this limit would require roughly
$\xi<\rL^{-(n+2)/n}$, or $\xi\lesssim0.09$ for these parameters,
provided the high-temperature-onset assumption continues to hold.
The example demonstrates why a small axion fraction cannot be inferred
from $\xi<1$ alone.

There is also a saturation regime. If twin confinement completes before
oscillations start and its susceptibility dominates, the mass is already
approximately constant. Then
\begin{equation}
 \Tosc\simeq
 \left[\frac{m_a(0)\Mbar}
 {3\sqrt{\pi^2g_*/90}}\right]^{1/2},
 \qquad
 \Omega_a\ \propto\ f_a^2\Theta_i^2\sqrt{m_a(0)}
 \label{eq:saturated-onset}
\end{equation}
at fixed $g_*$ and $g_{*s}$. Once this regime is reached, further
lowering $\xi$ does not continue the power-law suppression.
The two-temperature susceptibility and entropy evolution determine
where the transition occurs.

\subsection{Lattice-informed two-temperature evolution}
\label{sec:lattice-numerics}

The analytic power-law limit does not include the QCD crossover,
changes in the relativistic degrees of freedom, or the two baths'
contributions to the Friedmann equation. We therefore integrate the
small-angle homogeneous field using numerical QCD input. For the visible
sector, the data file accompanying this manuscript reproduces the
20 temperatures $100\,\mathrm{MeV}\le T_j\le3\,\mathrm{GeV}$ and
$\log_{10}[\chi(T_j)/\mathrm{fm}^{-4}]$ values from Table S7 of
Ref.~\cite{Borsanyi2016}. The conversion to natural units is
\begin{equation}
 \chi(T_j)=10^{q_j}\,(0.19732698\,\mathrm{GeV})^4,
 \qquad q_j=\log_{10}[\chi(T_j)/\mathrm{fm}^{-4}].
 \label{eq:lattice-units}
\end{equation}
We use shape-preserving cubic Hermite interpolation in $\ln T$
and $\ln\chi$. Below $100\,\mathrm{MeV}$ we use
the zero-temperature susceptibility $\chi_0=(5.70\times
10^{-3}\,\mathrm{GeV}^2)^2$, while above the last tabulated point we
continue with $\chi(T)\propto T^{-8.16}$. The latter is an explicitly
specified extrapolation, \emph{not} a lattice determination beyond
$3\,\mathrm{GeV}$. As in Eq.~\eqref{eq:finite-temperature-scaling}, the
\emph{uniform-spectrum} twin benchmark is
\begin{equation}
 \chi'(T')=r_\Lambda^4\chi(T'/r_\Lambda).
 \label{eq:lattice-twin-map}
\end{equation}
The distinct unchanged-quark-mass ($\kappa=1$) benchmark cannot be
calculated by this rescaling and is not included in the present scan.

\begin{figure}[t]
 \centering
 \includegraphics[width=0.97\columnwidth]{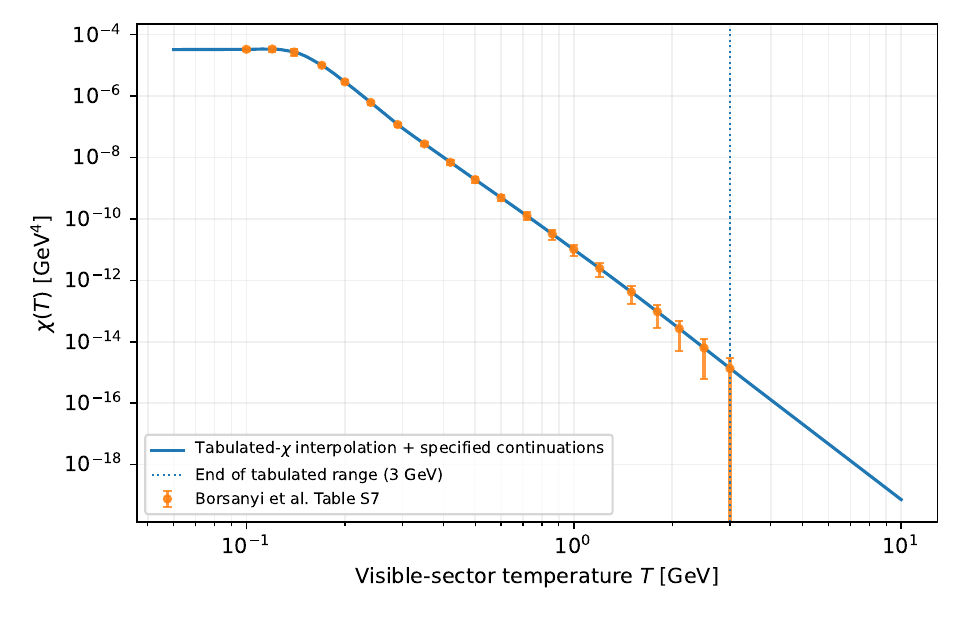}
 \caption{Topological susceptibility used in the numerical evolution.
 Points with quoted errors are the visible-QCD input of
 Ref.~\cite{Borsanyi2016}; the smooth line interpolates the tabulated
 data, with specified low- and high-temperature continuations. The
 dotted line marks the end of the published entries at $3\,\mathrm{GeV}$.
 The model for $\chi'$ is the conditional uniform rescaling in
 Eq.~\eqref{eq:lattice-twin-map}, not a separate twin lattice result.}
 \label{fig:lattice-chi}
\end{figure}

For the thermal background we sample the effective energy and entropy
degrees of freedom tabulated by Saikawa and Shirai
\cite{SaikawaShirai2018}, interpolating the supplied \emph{sparse}
subset in the data directory. Our particular uniformly rescaled
Standard Model proxy also adopts
\begin{equation}
 g_{*\rho}'(T')=g_{*\rho}(T'/r_\Lambda),\qquad
 g_{*s}'(T')=g_{*s}(T'/r_\Lambda).
 \label{eq:lattice-eos-map}
\end{equation}
This stronger \emph{homologous spectrum} assumption is independent of,
and not implied by, $\Lambda'_{\rm QCD}/\Lambda_{\rm QCD}=5$ alone.
It cannot substitute for a derived twin threshold spectrum or a
calculation of twin neutrino decoupling. In particular, the proxy uses
uniformly shifted thresholds rather than an explicitly calculated thermal
history for varying $\ralpha$. The numerical results below refer to the
specified homogeneous background; changing the twin electromagnetic
coupling requires re-evaluating thermal decoupling and, especially,
twin atomic recombination before making cosmological claims. We impose \emph{separate}
comoving-entropy conservation and define $x=\ln(A/A_R)$:
\begin{align}
 g_{*s}(T)T^3e^{3x}&=g_{*s}(T_R)T_R^3,\\
 g_{*s}'(T')T'^3e^{3x}&=g_{*s}'(T_R')T_R'^3,
 \label{eq:lattice-entropy}
\end{align}
with $T_R=100\,\mathrm{GeV}$ and $T_R'=\xi_RT_R$. At each $x$ we
compute $H$ from both radiation baths through
Eq.~\eqref{eq:hubble-two-temperature}, and evolve $\Theta$ and $d\Theta/dx$ using
Eq.~\eqref{eq:efold-evolution}. At sufficiently rapid oscillation we
extract the adiabatic number-to-visible-entropy ratio and multiply by
the \emph{zero-temperature} axion mass. Throughout this calculation
we set intersector energy transfer, thermal axions, anharmonicity,
late entropy production and a matter-dominated reheating stage to zero.

\begin{figure}[t]
 \centering
 \includegraphics[width=0.97\columnwidth]{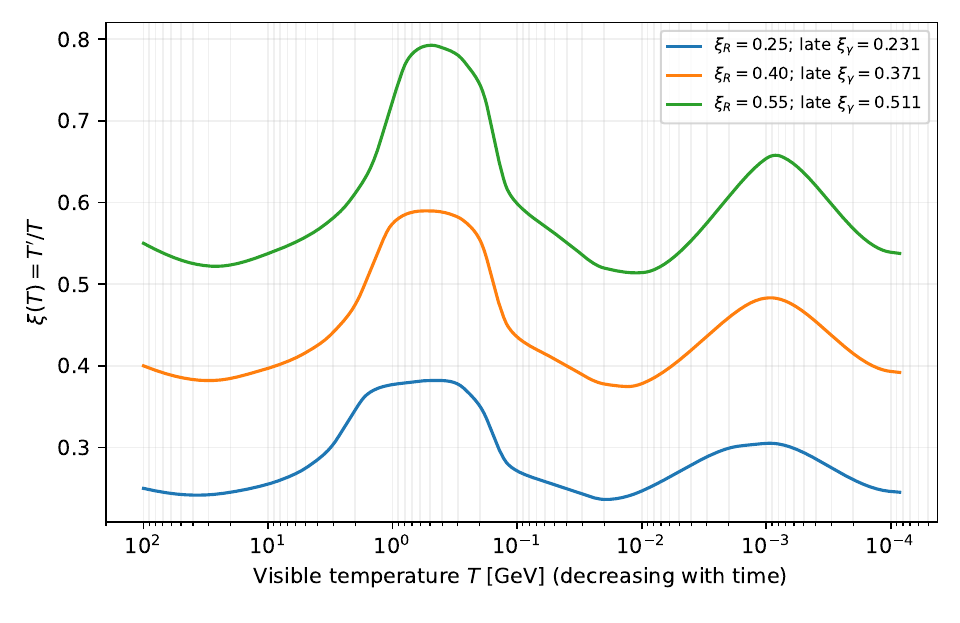}
 \caption{Evolution of the photon/plasma temperature ratio for three
 initial choices $\xi_R=T_R'/T_R$, under separate entropy conservation
 and the homologous equation of state in Eq.~\eqref{eq:lattice-eos-map}.
 The late photon ratios differ from their initial values because the
 particle thresholds in the two baths occur at different temperatures.
 The neutrino treatment is inherited from the rescaled Standard Model
 background and is not a twin-neutrino Boltzmann calculation.}
 \label{fig:temperature-ratio}
\end{figure}

For the reference choice
$r_\Lambda=5$, $f_a=10^{12}\,\mathrm{GeV}$,
$\Theta_i=0.10$, and $\xi_R=0.40$, the solution begins overdamped and
reaches the operational onset $m_a=3H$ at
$T_{\rm osc}\simeq8.90\,\mathrm{GeV}$ with
$T'_{\rm osc}\simeq3.56\,\mathrm{GeV}$. At the onset
$\xi_{\rm osc}\simeq0.3994$; after the modeled entropy releases,
$\xi_\gamma\simeq0.3711$. The resulting abundance is
\begin{equation}
 \Omega_a h^2\simeq0.00312,\qquad
 \fcos\simeq0.0260
 \quad(\Omega_{\rm DM}h^2=0.120).
 \label{eq:lattice-omega}
\end{equation}
Switching the numerical evolution to adiabatic-number conservation at
$m_a/H=200$ or $500$, instead of $300$, changes $\Omega_a h^2$ by
less than $0.1\%$; this is a convergence check on the integration, not
a systematic error bar. When the twin contribution to the axion
potential is switched off \emph{while holding the same two-bath
expansion background fixed}, the QCD reference abundance is
$0.00126$, so the calculated ratio is $2.48$. This need not equal the
$2.72$ result of the fixed-$g_*$, power-law toy model in
Sec.~\ref{sec:toy-numerics}; the input susceptibility and entropy
histories differ.

\begin{figure}[t]
 \centering
 \includegraphics[width=0.97\columnwidth]{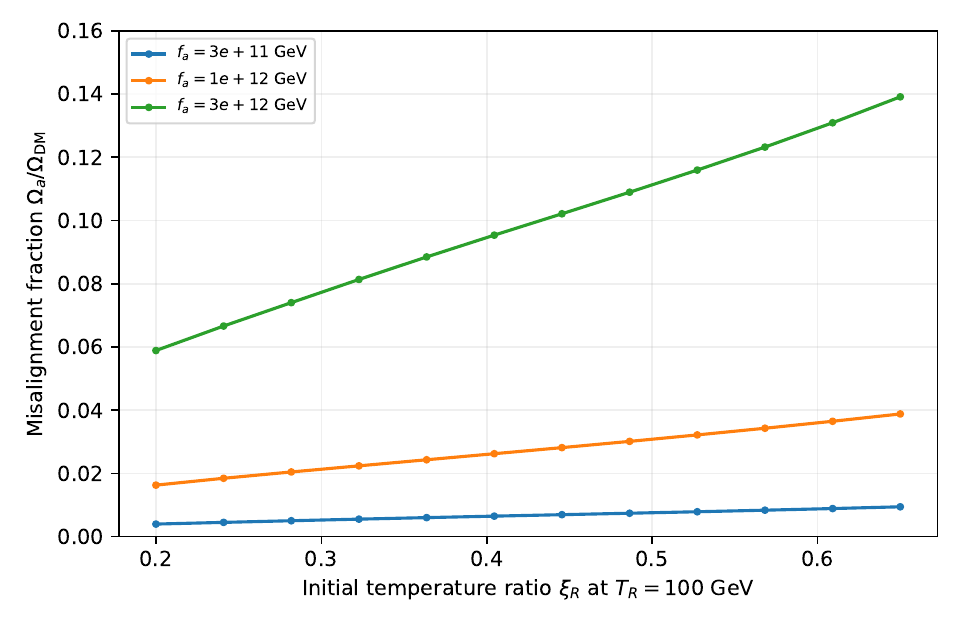}
 \caption{Misalignment fraction $\Omega_a/\Omega_{\rm DM}$ for
 $\Theta_i=0.1$, three decay constants, and 12 initial temperature
 ratios $0.20\le\xi_R\le0.65$. The background uses the tabulated QCD
 susceptibility and a sparse resampling of the Standard Model equation
 of state. All plotted points are conditional background abundances,
 \emph{not} cosmologically allowed regions or likelihood exclusions.}
 \label{fig:lattice-scan}
\end{figure}

The same separately conserved entropies give a late entropy ratio
$\zeta=s'/s\simeq0.05111$ for this benchmark. If the surviving light
twin radiation consists of a massless photon and three relativistic
twin neutrinos with the standard internal neutrino/photon temperature
ratio, Eq.~\eqref{eq:delta-neff} gives
$\Delta\Neff\simeq7.40\xi_\gamma^4\simeq0.140$. This is a
\emph{conditional} particle-content estimate and is not a CMB fit.
For $m_{B'}/m_p=5$, the required remainder of the dark-matter
budget, expressed as the twin baryon asymmetry per \emph{twin}
entropy relative to the visible yield, is
\begin{equation}
 \frac{Y_{B'}}{Y_B}=
 \frac{\Omega_{\rm DM}h^2-\Omega_a h^2}
 {\Omega_b h^2}\frac{m_p/m_{B'}}{\zeta}
 \simeq20.4\qquad(\Omega_b h^2=0.0224).
 \label{eq:lattice-asymmetry}
\end{equation}
The required asymmetry is an input, not a prediction of the axion anomaly.

\begin{table*}[t]
 \centering
 \caption{Numerical benchmark with tabulated visible-QCD input.
 All quantities are conditional on the homogeneous, uniformly rescaled,
 decoupled-bath assumptions specified in the text. The dark-radiation
 value additionally assumes a light twin photon and three light neutrinos.}
 \label{tab:lattice-benchmark}
 \begin{tabular*}{\textwidth}{@{\extracolsep{\fill}}p{0.35\textwidth}p{0.23\textwidth}p{0.32\textwidth}@{}}
 \toprule
 Quantity & Value & Interpretation \\
 \midrule
 $T_R,\;\xi_R,\;f_a,\;\Theta_i$
 & See text & Chosen initial conditions\\
 $T_{\rm osc},\;\xi_{\rm osc}$ & $8.90\,\GeV,\;0.3994$ & Numerical onset criterion $m_a=3H$\\
 $\xi_\gamma,\;\zeta$ & $0.3711,\;0.05111$ & Separate-entropy background\\
 $\Omega_a h^2,\;\fcos$ & $0.00312,\;0.0260$ & Homogeneous axion misalignment\\
 $\Omega_a/\Omega_{a,\mathrm{QCD}}$ & $2.48$ & Same radiation background, twin potential removed in reference\\
 $\Delta\Neff$ & $0.140$ & Assumed late relativistic twin content\\
 $Y_{B'}/Y_B$ & $20.4$ & Required budget for $m_{B'}/m_p=5$\\
 \bottomrule
 \end{tabular*}
\end{table*}

\begin{figure}[t]
 \centering
 \includegraphics[width=0.97\columnwidth]{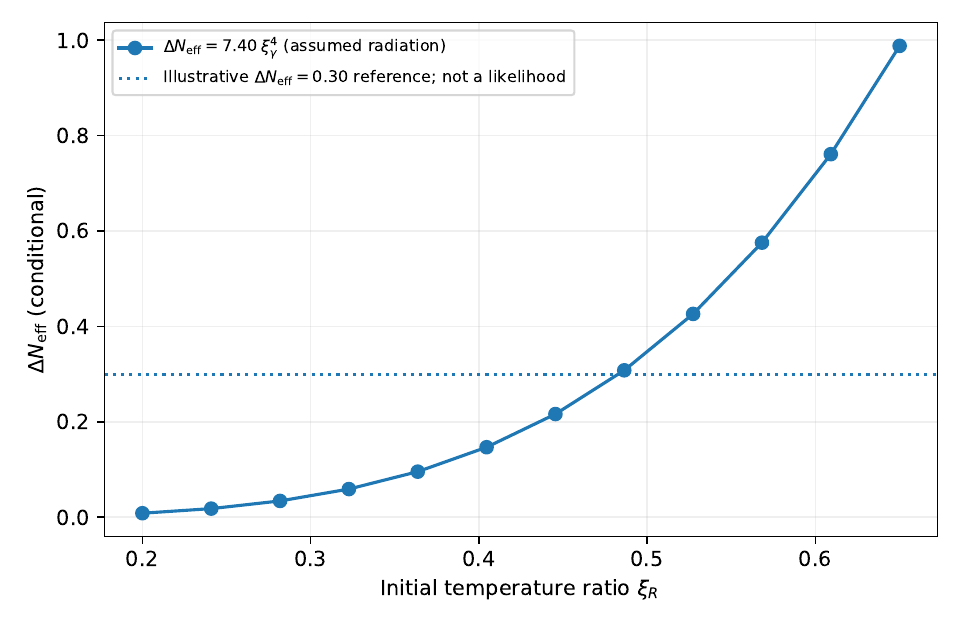}
 \caption{Illustrative extra-radiation budget from the same evolving
 temperature ratio, assuming one massless twin photon and three light
 twin neutrinos with the standard neutrino-to-photon temperature ratio.
 The horizontal $\Delta N_{\rm eff}=0.30$ guide is \emph{not} an
 observational likelihood. Interacting radiation and atomic dark matter
 require a joint perturbation calculation before model exclusion.}
 \label{fig:lattice-neff}
\end{figure}

\paragraph{Cosmological status.}
These calculations close the \emph{homogeneous} axion and baryon density
budget only after imposing a chosen baryon mass and asymmetry. They do
not evolve twin weak freeze-out, nuclear chemistry, recombination,
acoustic oscillations, ionization, velocity-dependent scattering, or the
linear CMB/matter perturbations. In particular, a dark-matter sector
that is almost entirely ionized or atomic baryons generally cannot be
treated as pressureless collisionless matter merely because
$\Delta N_{\rm eff}$ is small. A consistent full test will require a
specified twin spectrum, a recombination and perturbation solver, and
joint comparison with CMB anisotropies, Lyman-$\alpha$, and galactic
self-interaction observables \cite{CyrRacine2013,Bansal2022}.

\subsection{Independent analytic-limit check with a smooth toy susceptibility}
\label{sec:toy-numerics}

As an independent numerical regression test of the high-temperature
scaling without imposing an abrupt oscillation onset, we integrate the linearized homogeneous axion
field equation in the number of expansion e-folds,
$x=\ln(A/A_i)$, with $u=d\Theta/dx$:
\begin{equation}
 \frac{d\Theta}{dx}=u,\qquad
 \frac{du}{dx}=-\left(3+\frac{d\ln H}{dx}\right)u
 -\frac{m_a^2(T,T')}{H^2}\Theta.
 \label{eq:toy-field}
\end{equation}
For this controlled illustration we set
$g_*=g_{*s}=70$, $T'=\xi T$ with constant $\xi$, and assume
radiation domination with negligible twin radiation in $H$, no
intersector transfer, and no late entropy injection. These simplifying
assumptions are not a self-consistent precision thermal history.
We use the interpolating susceptibilities
\begin{equation}
 \chi_{\mathrm{toy}}(T)=\frac{\chi_0}{1+(T/T_c)^{2n}},
 \qquad
 \chi'_{\mathrm{toy}}(T')=r_\Lambda^4\,
 \chi_{\mathrm{toy}}(T'/r_\Lambda),
 \label{eq:toy-susceptibilities}
\end{equation}
with $T_c=150\,\mathrm{MeV}$ and $n=4$. The system is initialized
while $m_a/H\ll1$ and evolved into the adiabatic regime;
$Y_a=n_a/s$ is extracted from the oscillation-averaged energy density
and the zero-temperature mass is used in converting to $\Omega_a h^2$.
The functions in Eq.~\eqref{eq:toy-susceptibilities} are useful
continuity tests, not lattice-QCD fits \cite{Borsanyi2016}.

For $f_a=10^{12}\,\mathrm{GeV}$, $\Theta_i=0.1$,
$r_\Lambda=5$, and $\xi=0.4$, the ordinary-QCD reference gives
$\Omega_{a,\mathrm{QCD}}h^2\simeq0.00132$ while the twin-enhanced
case gives $\Omega_a h^2\simeq0.00358$. Their ratio is $2.7165$,
in agreement with the analytic high-temperature estimate $2.7166$.
The close agreement is a check of the stated limit, not a claim that
the complete twin-sector cosmology has been solved.

\begin{figure}[t]
\includegraphics[width=\columnwidth]{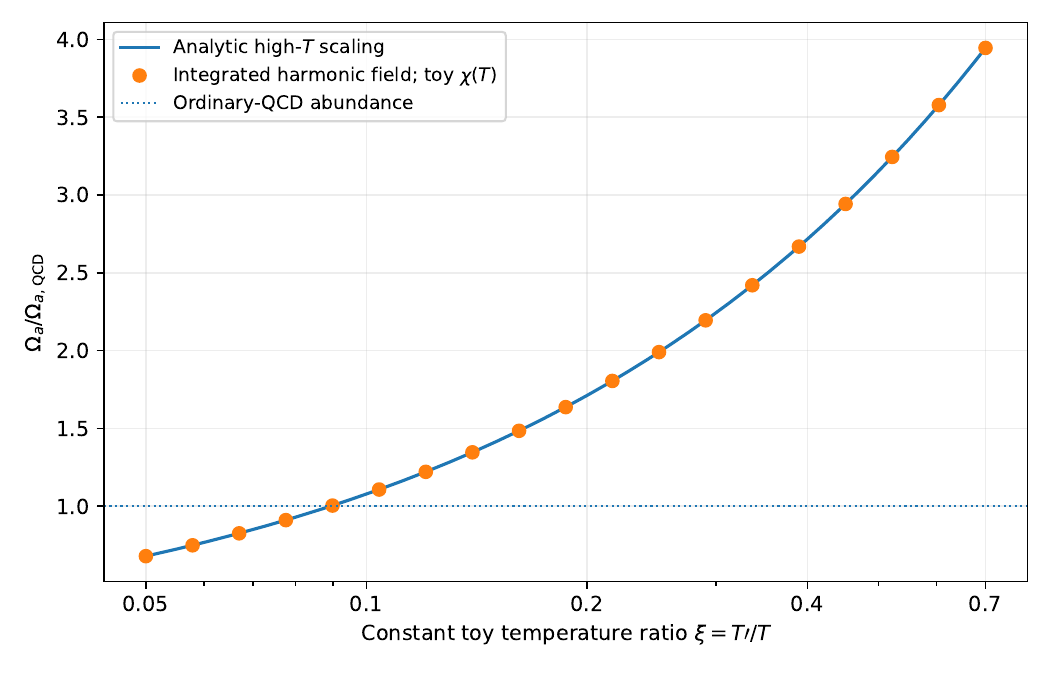}
\caption{A numerical test of the high-temperature abundance scaling.
Points show the ratio obtained by integrating the small-angle field
equation with Eq.~\eqref{eq:toy-susceptibilities}; the curve is
$(\sqrt{1+r_\Lambda^4}/r_\Lambda)\xi^{n/(n+2)}$ for
$r_\Lambda=5$ and $n=4$. All calculations use fixed $f_a$ and
$\Theta_i$, constant $g_*$ and $\xi$, and neglect twin radiation in
the expansion rate. This is a controlled toy benchmark, not an
observationally calibrated prediction.}
\label{fig:toy-relic}
\end{figure}

If we illustrate the abundance accounting with
$\Omega_{\mathrm{DM}}h^2=0.120$ and
$\Omega_bh^2=0.0224$, the above toy benchmark contains an axion
fraction $F_a^{\mathrm{cos}}\simeq0.030$. Assuming further that
$m_{B'}/m_p=5$ and the two late-time entropy degrees of freedom are
equal, $\zeta=s'/s=\xi^3=0.064$, the density budget would require
\begin{equation}
 \frac{Y_{B'}}{Y_B}=\frac{\Omega_{\mathrm{DM}}h^2-\Omega_a h^2}
 {\Omega_bh^2}\frac{m_p}{m_{B'}}\frac{1}{\zeta}
 \simeq16.2.
 \label{eq:toy-asymmetry}
\end{equation}
This is an algebraically consistent abundance assignment, not a
calculation of twin baryogenesis, recombination, or structure formation.

\section{Twin baryons as the dominant dark matter}
\label{sec:twin-baryons}

\subsection{The abundance coincidence and its entropy dependence}

For an asymmetry-dominated baryon population, neglecting binding-energy
corrections, the present densities obey
\begin{equation}
 \frac{\Omega_{B'}}{\Omega_b}
 =\frac{m_{B'}}{m_p}\frac{n_{B'}}{n_B}.
 \label{eq:baryon-abundance}
\end{equation}
Here $n_B$ and $n_{B'}$ denote the conserved baryon-number densities
remaining after symmetric annihilation. For mixtures of nuclei,
$m_{B'}$ should be interpreted as the mean mass per unit baryon number,
including any relevant binding and accompanying lepton contribution.
A fivefold baryon mass explains a fivefold energy density only when the
number-density ratio is approximately unity.

Define asymmetry yields using each sector's \emph{own} entropy:
\begin{equation}
 Y_B=\frac{n_B}{s},\qquad Y_{B'}=\frac{n_{B'}}{s'},\qquad
 \zeta=\frac{s'}s=\frac{g_{*s}'(T')}{g_{*s}(T)}\xi^3.
 \label{eq:baryon-yields}
\end{equation}
When both comoving entropies are conserved,
\begin{equation}
 \frac{\Omega_{B'}}{\Omega_b}
 =\frac{m_{B'}}{m_p}\frac{Y_{B'}}{Y_B}\zeta.
 \label{eq:asymmetry-abundance}
\end{equation}
Thus equal yields relative to each bath's entropy do \emph{not} imply
equal physical number densities in a colder twin sector.
For equal late entropy degrees of freedom and $\xi_\gamma=0.40$,
$\zeta=0.064$. Requiring $n_{B'}=n_B$ then demands
$Y_{B'}/Y_B\simeq15.6$. Equivalently, equal yields relative to a
\emph{common visible entropy} would imply equal number densities.
The distinction is a convention-dependent statement about ``equal
asymmetries,'' but its consequence for the density is physical.

Allowing an axion component and assuming no other dark matter gives
\begin{equation}
 \frac{Y_{B'}}{Y_B}=
 \frac{(1-\fcos)(\Omega_{\mathrm{DM}}/\Omega_b)}
      {(m_{B'}/m_p)\zeta}.
 \label{eq:required-asymmetry}
\end{equation}
The ratio $\Omega_{\mathrm{DM}}/\Omega_b\simeq5$ is the motivating
rounded benchmark; a data comparison should use the fitted ratio in
its chosen cosmological model \cite{Planck2018Parameters}.
Equation~\eqref{eq:required-asymmetry} links axion production, reheating,
and baryogenesis. These cannot all be fixed independently if the dark
sector is to saturate the observed matter density.

\subsection{Removal of the symmetric component}

A conserved asymmetry fixes the surviving baryon number but does not
automatically remove baryon--antibaryon pairs. Below confinement a
schematic annihilation equation is
\begin{equation}
 \dot n_{B'}^{\mathrm{part}}+3Hn_{B'}^{\mathrm{part}}
 =-\langle\sigma v\rangle
 \left[n_{B'}^{\mathrm{part}}n_{\bar B'}-
       n_{B',\mathrm{eq}}n_{\bar B',\mathrm{eq}}\right],
 \label{eq:baryon-annihilation}
\end{equation}
with the same annihilation term for $n_{\bar B'}$.
Their difference is conserved in the absence of baryon-number violation.
The condition for the asymmetry-dominated approximation is that the
residual symmetric component is small compared with that difference.
Strong annihilation is a motivation for this possibility, not a
substitute for computing it in the specified hadronic spectrum.
Quark-level evolution above confinement and hadronic annihilation below
it should be matched consistently.

\subsection{Stable baryons, nuclei, and atoms}
\label{sec:composition}

For light constituent quarks, the baryon mass is approximately set by
confinement, but the identity of the lightest baryon depends on isospin
and electromagnetic effects. A schematic separation is
\begin{equation}
 m_{n'}-m_{p'}\simeq
 A_I(m_d'-m_u')-B_{\mathrm{EM}}\alpha'_{\mathrm{EM}}
                     \Lambda'_{\mathrm{QCD}},
 \label{eq:baryon-splitting}
\end{equation}
with dimensionless coefficients determined by the hadronic theory.
The competition of these contributions is established in visible
QCD plus electromagnetism \cite{Borsanyi2015}.
With uniform scaling and fixed dimensionless couplings, both terms
scale together. With unchanged light-quark masses but enhanced
confinement, their relative importance changes; the stable twin baryon
cannot simply be assumed to remain the proton.

Ignoring neutrino masses, free-neutron beta decay requires
$m_{n'}-m_{p'}>m_{e'}$, while free-proton beta-plus decay requires
$m_{p'}-m_{n'}>m_{e'}$. Bound-state electron capture introduces a separate
condition involving the atomic binding energy. Nuclear binding can
change the surviving composition further. A neutron-dominated sector,
an ionized proton--electron plasma, and neutral twin atoms have very
different scattering and cooling properties even at the same
$\Omega_{B'}$.

For a hydrogen-like twin atom,
\begin{equation}
 a_0'=\frac{1}{\alpha'_{\mathrm{EM}}\mu'},\qquad
 E_B'=\frac12(\alpha'_{\mathrm{EM}})^2\mu',\qquad
 \mu'=\frac{m_{e'}m_{p'}}{m_{e'}+m_{p'}}.
 \label{eq:atomic-scales}
\end{equation}
If the twin proton and electron are each five times heavier,
$m_{p'}=5m_p$ and $m_{e'}=5m_e$, these scales become $a_0'=a_0/(5\ralpha)$ and
$E_B'=5\ralpha^2 E_B$. The value of $\ralpha$ must be obtained from
ultraviolet and threshold matching; $\ralpha=1$ is used only for
the following numerical size estimate. Even if the twin unification
scale differs from the visible one, the hierarchy alone does not
uniquely determine $\alpha'_{\mathrm{EM}}$.
Recombination, residual ionization, and kinetic decoupling then depend
on $m_{e'}$, the radiation temperature, and the twin baryon-to-photon
ratio. The binding energy alone is not the recombination temperature.
Atomic dark matter can exhibit dark acoustic oscillations and
velocity-dependent self-interactions
\cite{CyrRacine2013,Bansal2022}. Nuclear abundances also depend on twin
weak freeze-out and nuclear reaction rates, not just rescaled baryon
masses.

For purely hadronic scattering, uniform scaling suggests a dimensional
estimate $\sigma_{B'B'}\sim\Lambda_{\mathrm{QCD}}'^{-2}$ and hence
$\sigma_{B'B'}/m_{B'}\propto\rL^{-3}$ at fixed dimensionless kinematics.
This estimate is not a prediction near nuclear resonances and says
nothing about potentially much larger atomic cross sections.
Astrophysical self-interaction tests instead constrain a velocity- and
composition-dependent \emph{transport} cross section per unit mass;
what is relevant for dwarf galaxies may differ from what is relevant
for galaxy clusters. A neutron-dominated twin sector could be governed
primarily by short-range hadronic scattering, whereas charged or neutral
twin atoms can exhibit collisional, radiative, and acoustic dynamics.
Neither the scaling estimate nor the axion abundance calculation supplies
a predicted momentum-transfer cross section. Testing dominant twin-baryon
dark matter requires the cross section at the relevant halo velocities,
plus its ionized, atomic, nuclear, and dissipative fractions
\cite{CyrRacine2013,Bansal2022,Barron2026Atomic,Yuan2026DAO}.

An elementary atomic size estimate emphasizes the required next step.
For the illustrative choice $m_{p'}=5m_p$, $m_{e'}=5m_e$,
and $\ralpha=1$, the hydrogen-like binding energy is approximately
$E_B'\simeq68\,\mathrm{eV}$ and
$a_0'\simeq1.06\times10^{-9}\,\mathrm{cm}$. At general $\ralpha$,
these become $E_B'\simeq68\ralpha^2\,\mathrm{eV}$ and
$a_0'\simeq1.06\times10^{-9}\ralpha^{-1}\,\mathrm{cm}$.
The ratio of the \emph{geometric} area $\pi a_0'^2$ to the assumed
atomic mass is of order $4\times10^5\ralpha^{-2}\,
\mathrm{cm^2\,g^{-1}}$.
This number is only a size-scale diagnostic: the momentum-transfer
cross section is sensitive to collision velocity, quantum molecular
potentials, ionization, and the actual atomic composition. In
particular it must \emph{not} be interpreted as a predicted
self-interaction cross section or a direct exclusion. With a
$\gtrsim 97\%$ twin-baryon fraction, computing twin recombination
and dark acoustic damping is necessary before declaring the example
compatible with cosmological data.

For an unbroken twin electromagnetic group, a charged lightest
baryon may bind with twin electrons. Such a dark component need not
behave as pressureless collisionless matter, even if the homogeneous
abundance budget closes. Dark acoustic oscillations and dissipative
cooling are constrained by CMB, galaxy-formation, and small-scale-structure
measurements \cite{CyrRacine2013,Barron2026Atomic,Yuan2026DAO}.
A model-level constraint requires the stable twin hadron spectrum,
recombination and residual ionization histories, velocity-dependent
transport cross sections, and the resulting linear matter transfer
function. Constraints inferred in simplified atomic-dark-matter models
cannot be adopted as automatic exclusions of the present F-theory-motivated
benchmark without matching the particle content and interaction history.

\section{Inflationary initial conditions and topological defects}
\label{sec:inflation}

If the Peccei--Quinn symmetry is broken before inflation and never
restored, our observable patch may have a single initial angle.
For a light, canonically normalized axion during inflation and an
unchanged decay constant,
\begin{equation}
 \mathcal P_{\delta\Theta}=\left(\frac{H_I}{2\pi f_a}\right)^2.
 \label{eq:inflation-fluctuation}
\end{equation}
In the harmonic misalignment regime, with the other dark matter
adiabatic and uncorrelated with the axion fluctuation,
\begin{equation}
 \mathcal P_S\simeq(\fcos)^2
 \left(\frac{H_I}{\pi f_a\Theta_i}\right)^2,
 \qquad
 \beta_{\mathrm{iso}}=
 \frac{\mathcal P_S}{\mathcal P_{\mathcal R}+\mathcal P_S}.
 \label{eq:isocurvature}
\end{equation}
For a chosen upper limit $\beta_{\max}$ this implies
\begin{equation}
 H_I\lesssim\frac{\pi f_a|\Theta_i|}{\fcos}
 \left[\frac{\beta_{\max}}{1-\beta_{\max}}
                  \mathcal P_{\mathcal R}\right]^{1/2}.
 \label{eq:isocurvature-bound}
\end{equation}
Planck 2018 supplies reference likelihoods for specific isocurvature
models \cite{Planck2018Inflation}. These linearized expressions also
assume a classical displacement larger than its fluctuations,
$|\Theta_i|\gg H_I/(2\pi f_a)$; the fluctuation-dominated regime
requires a statistical treatment. The formula should not be used
unchanged for a heavy inflationary axion, a different inflationary
kinetic normalization, a large-angle abundance, or correlated
twin-baryon isocurvature. In particular, selecting a baryogenesis
mechanism can introduce an additional fluctuation rather than merely
fixing the background abundance.

If the symmetry is instead broken after inflation, averaging
$\Theta_i^2$ over patches is not a complete abundance calculation:
strings and domain walls can radiate axions. To identify the wall
number, restore the fundamental periodic variable
$\vartheta=a/f_{\mathrm{PQ}}$ with period $2\pi$ and integer color
anomalies $N$ and $N'$. For aligned potentials with no additional
identifications, the exact common periodicity gives
\begin{equation}
 N_{\mathrm{DW}}=\gcd(|N|,|N'|).
 \label{eq:domain-wall-number}
\end{equation}
In particular, $N=N'$ leaves $N_{\mathrm{DW}}=|N|$, not $2|N|$.
The normalization $f_a=f_{\mathrm{PQ}}/N$ used in the color coupling
hides this integer and must not be taken as proof that
$N_{\mathrm{DW}}=1$. Additional gauge identifications or explicit
breaking can change the counting. In the equal-anomaly case, two
confinement contributions do not by themselves remove a stable
$N_{\mathrm{DW}}>1$ wall problem; a proposed bias must also satisfy
the strong-CP condition in Eq.~\eqref{eq:quality-bound}.

\section{From benchmark relations to a testable model}
\label{sec:strategy}

The preceding equations define a sequence of calculations rather than
a single exclusion plot. First, the ultraviolet matching fixes the
light axion direction, its anomalies, and the twin spectrum. The
matched spectrum fixes the susceptibilities and the identity of stable
matter. The reheating and entropy history then fixes $T'(T)$ and the
background expansion used in the axion and baryon calculations.
Finally, the resulting matter fractions and interactions determine
which cosmological and experimental likelihoods apply.

A useful first comparison keeps the two $\rL=5$ susceptibility
benchmarks but varies $f_a$, the initial angle, and the temperature ratio.
The $\kappa=5$ case can use a consistently rescaled visible-QCD thermal
input; the $\kappa=1$ case cannot use that rescaling without an additional
approximation. A small observed axion signal could be explained by a
small intrinsic coupling, a small local axion density, or both, but the
first is a particle-physics output and the second is a population and
halo output. They should be displayed separately.

Table~\ref{tab:cosmo-checks} distinguishes earlier analytic or toy
consistency checks from the single conditional numerical background
benchmark in Table~\ref{tab:lattice-benchmark}. Neither table is a
fitted, fully viable cosmology. In particular, $\xi$ at oscillation
and $\xi_\gamma$ at late times need not agree.

\begin{table*}[t]
 \centering
 \caption{Conditional consistency checks for the uniformly rescaled
 $\rL=5$ benchmark. No row asserts that all constraints have been met.}
 \label{tab:cosmo-checks}
 \begin{tabular*}{\textwidth}{@{\extracolsep{\fill}}p{0.27\textwidth}p{0.36\textwidth}p{0.27\textwidth}@{}}
 \toprule
 Input & Relation tested & Consequence\\
 \midrule
 $\chi'/\chi=625$, $E/N=8/3$
 & Eq.~\eqref{eq:benchmark-line}
 & $\Ra=25.02$; fixed intrinsic slope\\[3pt]
 Three relativistic twin neutrinos; $\xi_\gamma=0.40$
 & Eq.~\eqref{eq:delta-neff}, with standard neutrino temperature ratio
 & $\Delta\Neff\simeq0.19$\\[3pt]
 Equal late entropy degrees of freedom; $n_{B'}=n_B$
 & Eq.~\eqref{eq:asymmetry-abundance}, with $\xi_\gamma=0.40$
 & $Y_{B'}/Y_B\simeq15.6$\\[3pt]
 $n=4$, $\xi(\Tosc)=0.40$; high-temperature onset
 & Eq.~\eqref{eq:relic-ratio}, equal $f_a$ and $\Theta_i$
 & $\Omega_a/\Omega_{a,\mathrm{QCD}}\simeq2.7$\\[3pt]
 $f_a=10^{12}\,\GeV$, $\Theta_i=0.1$; lattice-informed, separately adiabatic baths
 & Eqs.~\eqref{eq:lattice-omega}--\eqref{eq:lattice-asymmetry}
 & $\fcos\simeq0.026$; \newline $Y_{B'}/Y_B\simeq20.4$ (conditional)\\[3pt]
 Small $\bar\theta-\bar\theta'$
 & Eq.~\eqref{eq:theta-mismatch}
 & Almost no suppression of the mismatch\\
 \bottomrule
 \end{tabular*}
\end{table*}

The minimal population requirement, assuming no other dark matter, is
\begin{equation}
 \Omega_a+\Omega_{B'}=\Omega_{\mathrm{DM}},\qquad
 0\leq\fcos\leq1.
 \label{eq:population-closure}
\end{equation}
It must be imposed together with radiation, isocurvature, stability,
and structure-formation tests, not before calculating them.
The local fraction $\flocal$ appearing in Eq.~\eqref{eq:limit-rescaling}
can then be related to $\fcos$ only under a specified halo model.
Dissipative twin matter makes that last step especially nontrivial.

\section{Conclusion}
\label{sec:conclusion}
A universal axion coupled to visible and twin QCD can be heavier at
fixed decay constant, provided the two physical strong-CP minima are
aligned and the canonically normalized axion anomalies are correctly
matched. The physical mass follows from the sum of the two
susceptibilities. In the uniformly rescaled $r_\Lambda=5$ benchmark
$R_a=\sqrt{626}\simeq25.02$, and assuming $E/N=8/3$ gives the intrinsic
visible-photon relation
$|g_{a\gamma\gamma}|\simeq6.1\times10^{-12}
(m_a/\mathrm{eV})\,\mathrm{GeV}^{-1}$. It agrees with the marked
benchmark in the supplied axion-coupling plot. The visible-photon
coupling and a local dark-matter-signal rescaling must remain separate.

Beyond this particle-physics locus, we have advanced the cosmology from
an idealized analytic estimate to a reproducible numerical
\emph{background} calculation. We interpolate published
finite-temperature visible-QCD topological-susceptibility values,
combine them with a sampled Standard Model equation of state, track
separate visible and twin entropy release, and include both radiation
baths in $H$. Under a uniform full-spectrum rescaling and no
intersector energy transfer, a benchmark with
$f_a=10^{12}\,\mathrm{GeV}$, $\Theta_i=0.10$ and $\xi_R=0.40$
produces $\Omega_a h^2\simeq0.00312$, about $2.6\%$ of an illustrative
$\Omega_{\rm DM}h^2=0.120$. The evolved late photon ratio is
$\xi_\gamma\simeq0.371$, in contrast to the initial $0.40$. A
conditionally light twin photon and three light twin neutrinos would
then yield $\Delta N_{\rm eff}\simeq0.14$. To fill the remaining
abundance with five-times-heavier twin baryons would require
$Y_{B'}/Y_B\simeq20.4$ when the asymmetries are normalized to each
sector's entropy.

These numbers are \emph{not} a self-contained viable mixed-dark-matter
model. The entropy evolution uses a sparse, interpolated Standard Model
equation of state and a homologous twin proxy; the susceptibility is
extrapolated above $3\,\mathrm{GeV}$; and we have not included axion
transfer between baths, thermal axions, nonlinear misalignment,
twin baryogenesis, recombination, or matter perturbations. In
particular, a dark sector dominated by stable charged or atomic twin
baryons requires detailed checks of dark acoustic oscillations,
dissipative dynamics, and velocity-dependent self-interactions.
Preliminary equality of visible and twin ultraviolet Yukawa couplings,
when established, would still require Higgs-VEV and threshold matching;
likewise the ratio $\ralpha$ must be determined before the atomic
benchmark becomes a prediction. These requirements are in addition to
the strong-CP and axion-normalization consistency tests.
The accompanying tabulated inputs and integration code make the
conditional background calculation reproducible and provide a
starting point for those necessary model-dependent tests.

\begin{acknowledgments}
We thank Gordon Kane and Bobby Acharya for useful conversations. This research was supported in part by U.S. Department of Energy grant DE-FG02-85ER40237, DE-GR102719 and the Simons Center. 
\end{acknowledgments}

\appendix
\section{Numerical evolution: a reproducible prescription}
\label{app:numerics}

This appendix specifies the evolution equations used in both the
lattice-informed background calculation in Sec.~\ref{sec:lattice-numerics}
and the independent smooth-toy regression test in
Sec.~\ref{sec:toy-numerics}. The lattice-informed calculation uses a
sparse, interpolated Standard Model equation of state and the stated
homologous twin-sector assumptions; neither calculation includes
linear cosmological perturbations or a joint observational fit.
The natural independent variable is the expansion e-fold
$x=\ln(A/A_i)$. In a separately adiabatic visible bath,
\begin{equation}
 \frac{\dd\ln T}{\dd x}=
 -\left[1+\frac13\frac{\dd\ln g_{*s}}{\dd\ln T}\right]^{-1},
 \label{eq:temperature-evolution}
\end{equation}
and the twin temperature follows from its own entropy equation or
Eq.~\eqref{eq:xi-entropy}. If intersector transfer or a decaying matter
component is important, the corresponding energy equations replace
this adiabatic relation.

Write $u=\dd\Theta/\dd x$. The axion equation becomes
\begin{align}
 \frac{\dd\Theta}{\dd x}&=u,\\
 \frac{\dd u}{\dd x}&=
 -\left(3+\frac{\dd\ln H}{\dd x}\right)u
 -\frac{1}{f_a^2H^2}\frac{\partial V}{\partial\Theta}.
 \label{eq:efold-evolution}
\end{align}
For radiation with constant degrees of freedom,
$\dd\ln H/\dd x=-2$, providing a useful check of the damping term.
Initial conditions $\Theta=\Theta_i$ and $u=0$ can be imposed sufficiently
early that $m_a/H\ll1$, provided the field was initially overdamped.
One must not impose this condition after the onset of motion merely
because the visible sector has not yet confined.

For testing an integrator, a smooth toy susceptibility is
\begin{equation}
 \chi_{\mathrm{toy}}(T)=
 \frac{\chi_0}{1+(T/T_c)^{2n}},\qquad
 \chi'_{\mathrm{toy}}(T')=
 \rL^4\chi_{\mathrm{toy}}(T'/\rL).
 \label{eq:toy-susceptibility}
\end{equation}
It interpolates between the assumed limiting powers, but is not a
lattice fit. Choosing it and a cosine potential is a numerical
benchmark that should be labeled as such. The changed-quark-mass
case needs a separate thermal input. For publication-quality
abundances, one should propagate uncertainties from the susceptibility,
equation of state, and nonharmonic potential rather than varying only
$\Theta_i$.

After many adiabatic oscillations, extract
\begin{equation}
 Y_a=\frac{\left\langle
 \tfrac12 f_a^2 H^2u^2+V(\Theta;T,T')-V_{\min}(T,T')
 \right\rangle}{m_a(T,T')s(T)}.
 \label{eq:numerical-yield}
\end{equation}
The average must extend over several oscillations in a regime where
the potential is effectively harmonic and $Y_a$ has stabilized.
Convergence should be checked against the integration tolerance,
initial time, averaging interval, and the time of the switch to
adiabatic number conservation.

\paragraph{Data and reproducibility.}
The complete project contains the original 20-row susceptibility
values and quoted one-sigma logarithmic uncertainties in
\path{data/borsanyi2016_table_S7.csv}, and a 29-row sparse sample
of the published Saikawa--Shirai degrees of freedom in
\path{data/eos2018_sampled.csv}. These are finite-temperature
visible-sector inputs, not derived twin datasets. The file
\path{code/lattice_two_temperature_scan.py} implements the
monotone log-log interpolation, separate-entropy inversion, Hubble
rate, harmonic field evolution, adiabatic number conversion, and
scans underlying Figs.~\ref{fig:lattice-chi}--\ref{fig:lattice-neff}.
The illustrative numerical conversion uses
$s_0=2891.2\,\mathrm{cm}^{-3}$ and
$\rho_c/h^2=1.05375\times10^{-5}\,\mathrm{GeV\,cm^{-3}}$.
Both the external table interpolation and all additional cosmological
assumptions must be varied for a credible precision uncertainty
estimate. Overleaf reads the precomputed PDF figures, so compilation
does not require Python. A README documents how to reproduce the
results locally.

Three limiting checks are particularly useful. Removing the twin
susceptibility must recover the single-sector evolution. Identical
sectors at equal temperature give $m_a^2(T)=2\chi(T)/f_a^2$ and require
both baths in $H$. Finally, a twin-dominated, power-law onset with
negligible twin radiation must approach Eq.~\eqref{eq:relic-ratio};
a sufficiently early twin crossover must instead approach
Eq.~\eqref{eq:saturated-onset}. These tests separate errors in the
potential, the background, and the final mass conversion.

\section{Finite-angle alignment in a cosine approximation}
\label{app:alignment}

The small-mismatch relation in the main text does not assume cosine
potentials. For illustration only, take
\begin{equation}
 V_{\cos}(\Theta)=-\chi\cos\Theta-\chi'\cos(\Theta+\Delta).
 \label{eq:cosine-two-sector}
\end{equation}
Define the complex quantity
$Z=\chi+\chi'e^{i\Delta}$. Then
$V_{\cos}=-|Z|\cos(\Theta+\arg Z)$, so the stable minimum and curvature are
\begin{equation}
 \Theta_{\min}=-\arg Z\quad(\mathrm{mod}\ 2\pi),\qquad
 m_a^2 f_a^2=|Z|
 =\sqrt{\chi^2+\chi'^2+2\chi\chi'\cos\Delta}.
 \label{eq:cosine-exact}
\end{equation}
Expanding at small $\Delta$ reproduces
$\Theta_{\min}=-\chi'\Delta/(\chi+\chi')$.
This example also illustrates why the aligned susceptibility sum cannot
be applied unchanged at arbitrary relative phase. For large $\Delta$,
the visible strong-CP angle is generally large even when the axion is
heavy. The full chiral or nonperturbative potential, rather than
Eq.~\eqref{eq:cosine-exact}, is required for quantitative large-angle
claims in QCD.

\bibliography{references}

@article{ClemensRaby2020,
  author        = {Clemens, C. Herbert and Raby, Stuart},
  title         = {Relative Scales of the {GUT} and Twin Sectors in an {F}-theory model},
  journal       = {JHEP},
  volume        = {2020},
  number        = {04},
  pages         = {004},
  year          = {2020},
  doi           = {10.1007/JHEP04(2020)004},
  eprint        = {2001.10047},
  archivePrefix = {arXiv},
  primaryClass  = {hep-th},
  note          = {See revised version v2}
}

@article{Berezhiani2001,
  author        = {Berezhiani, Zurab and Gianfagna, Leonida and Giannotti, Maurizio},
  title         = {Strong {CP} problem and mirror world: the {Weinberg--Wilczek} axion revisited},
  journal       = {Phys. Lett. B},
  volume        = {500},
  pages         = {286--296},
  year          = {2001},
  doi           = {10.1016/S0370-2693(00)01392-7},
  eprint        = {hep-ph/0009290},
  archivePrefix = {arXiv}
}

@article{Dunsky2023,
  author        = {Dunsky, David I. and Hall, Lawrence J. and Harigaya, Keisuke},
  title         = {A Heavy {QCD} Axion and the Mirror World},
  journal       = {JHEP},
  volume        = {2024},
  number        = {02},
  pages         = {212},
  year          = {2024},
  doi           = {10.1007/JHEP02(2024)212},
  eprint        = {2302.04274},
  archivePrefix = {arXiv},
  primaryClass  = {hep-ph}
}

@article{GrilliDiCortona2016,
  author        = {{Grilli di Cortona}, Giovanni and Hardy, Edward and {Pardo Vega}, Javier and Villadoro, Giovanni},
  title         = {The {QCD} axion, precisely},
  journal       = {JHEP},
  volume        = {2016},
  number        = {01},
  pages         = {034},
  year          = {2016},
  doi           = {10.1007/JHEP01(2016)034},
  eprint        = {1511.02867},
  archivePrefix = {arXiv},
  primaryClass  = {hep-ph}
}

@misc{OHareAxionLimits,
  author       = {O'Hare, Ciaran A. J.},
  title        = {{AxionLimits}: FIPS Workshop 2022
                  Axion--Photon Coupling Compilation},
  year         = {2022},
  howpublished = {\url{https://github.com/cajohare/AxionLimits/blob/master/docs/fips.md}},
  note         = {FIPS Workshop 2022 figure.
                  AxionLimits repository archived at
                  DOI: 10.5281/zenodo.3932430}
}

@article{Borsanyi2016,
  author        = {Borsanyi, S. and others},
  title         = {Calculation of the axion mass based on high-temperature lattice quantum chromodynamics},
  journal       = {Nature},
  volume        = {539},
  pages         = {69--71},
  year          = {2016},
  doi           = {10.1038/nature20115},
  eprint        = {1606.07494},
  archivePrefix = {arXiv},
  primaryClass  = {hep-lat}
}

@article{Giannotti2005,
  author        = {Giannotti, Maurizio},
  title         = {Mirror World and Axion: Relaxing Cosmological Bounds},
  journal       = {Int. J. Mod. Phys. A},
  volume        = {20},
  pages         = {2454--2458},
  year          = {2005},
  eprint        = {astro-ph/0504636},
  archivePrefix = {arXiv}
}

@article{Planck2018Parameters,
  author        = {Aghanim, N. and others},
  collaboration = {Planck},
  title         = {{Planck} 2018 results. {VI}. Cosmological parameters},
  journal       = {Astron. Astrophys.},
  volume        = {641},
  pages         = {A6},
  year          = {2020},
  doi           = {10.1051/0004-6361/201833910},
  eprint        = {1807.06209},
  archivePrefix = {arXiv},
  primaryClass  = {astro-ph.CO}
}

@article{Planck2018Inflation,
  author        = {Akrami, Y. and others},
  collaboration = {Planck},
  title         = {{Planck} 2018 results. {X}. Constraints on inflation},
  journal       = {Astron. Astrophys.},
  volume        = {641},
  pages         = {A10},
  year          = {2020},
  doi           = {10.1051/0004-6361/201833887},
  eprint        = {1807.06211},
  archivePrefix = {arXiv},
  primaryClass  = {astro-ph.CO}
}

@article{Abel2020,
  author        = {Abel, C. and others},
  title         = {Measurement of the permanent electric dipole moment of the neutron},
  journal       = {Phys. Rev. Lett.},
  volume        = {124},
  pages         = {081803},
  year          = {2020},
  doi           = {10.1103/PhysRevLett.124.081803},
  eprint        = {2001.11966},
  archivePrefix = {arXiv},
  primaryClass  = {hep-ex}
}

@article{Hook2020,
  author        = {Hook, Anson and Kumar, Soubhik and Liu, Zhen and Sundrum, Raman},
  title         = {The High Quality {QCD} Axion and the {LHC}},
  journal       = {Phys. Rev. Lett.},
  volume        = {124},
  pages         = {221801},
  year          = {2020},
  doi           = {10.1103/PhysRevLett.124.221801},
  eprint        = {1911.12364},
  archivePrefix = {arXiv},
  primaryClass  = {hep-ph}
}

@article{Bansal2022,
  author        = {Bansal, Saurabh and Kim, Jeong Han and Kolda, Christopher and Low, Matthew and Tsai, Yuhsin},
  title         = {Mirror Twin Higgs Cosmology: Constraints and a Possible Resolution to the {$H_0$} and {$S_8$} Tensions},
  journal       = {JHEP},
  volume        = {2022},
  number        = {05},
  pages         = {050},
  year          = {2022},
  doi           = {10.1007/JHEP05(2022)050},
  eprint        = {2110.04317},
  archivePrefix = {arXiv},
  primaryClass  = {hep-ph}
}

@article{Borsanyi2015,
  author        = {Borsanyi, S. and others},
  title         = {Ab initio calculation of the neutron-proton mass difference},
  journal       = {Science},
  volume        = {347},
  pages         = {1452--1455},
  year          = {2015},
  doi           = {10.1126/science.1257050},
  eprint        = {1406.4088},
  archivePrefix = {arXiv},
  primaryClass  = {hep-lat}
}

@article{CyrRacine2013,
  author        = {Cyr-Racine, Francis-Yan and Sigurdson, Kris},
  title         = {The Cosmology of Atomic Dark Matter},
  journal       = {Phys. Rev. D},
  volume        = {87},
  pages         = {103515},
  year          = {2013},
  doi           = {10.1103/PhysRevD.87.103515},
  eprint        = {1209.5752},
  archivePrefix = {arXiv},
  primaryClass  = {astro-ph.CO}
}

@article{Clemens:2019dts,
  author        = {Clemens, Herbert and Raby, Stuart},
  title         = {Heterotic-{F}-theory Duality with
                   {Wilson} Line Symmetry-breaking},
  journal       = {JHEP},
  volume        = {2019},
  number        = {12},
  pages         = {016},
  year          = {2019},
  doi           = {10.1007/JHEP12(2019)016},
  eprint        = {1908.01913},
  archivePrefix = {arXiv},
  primaryClass  = {hep-th}
}

@article{Calabrese2025ACT,
  author = {Erminia Calabrese and others},
  title = {The Atacama Cosmology Telescope: DR6 constraints on extended cosmological models},
  journal = {Journal of Cosmology and Astroparticle Physics},
  volume = {2025},
  number = {11},
  pages = {063},
  year = {2025},
  doi = {10.1088/1475-7516/2025/11/063},
  eprint = {2503.14454},
  archivePrefix = {arXiv}
}

@article{Barron2026Atomic,
  author = {Jared Barron and Rouven Essig and Megan H. McDuffie and
            Jes{\'u}s P{\'e}rez-R{\'i}os and Gregory Suczewski},
  title = {Pushing the limits of atomic dark matter: First-principles
           recombination rates and cosmological constraints},
  journal = {Journal of Cosmology and Astroparticle Physics},
  volume = {2026},
  number = {07},
  pages = {090},
  year = {2026},
  doi = {10.1088/1475-7516/2026/07/090},
  eprint = {2602.10197},
  archivePrefix = {arXiv}
}

@misc{Yuan2026DAO,
  author = {Zhihan Yuan and Caleb Gemmell and Keir K. Rogers and
            Jared Barron and Sandip Roy and David Curtin and Norman Murray},
  title = {Strongest constraints on dark acoustic oscillations from the Lyman-alpha forest},
  year = {2026},
  eprint = {2606.28482},
  archivePrefix = {arXiv},
  primaryClass = {astro-ph.CO}
}

@misc{Trendafilova2026Interacting,
  author = {Cynthia Trendafilova and Fei Ge and Ali Rida Khalife and
            Christian L. Reichardt},
  title = {Cosmological Constraints on Interacting Dark Matter with SPT-3G D1},
  year = {2026},
  eprint = {2610.37604},
  archivePrefix = {arXiv},
  primaryClass = {astro-ph.CO}
}

@article{SaikawaShirai2018,
  author = {Saikawa, Ken'ichi and Shirai, Satoshi},
  title = {Primordial gravitational waves, precisely: the role of thermodynamics in the Standard Model},
  journal = {JCAP},
  volume = {2018},
  number = {05},
  pages = {035},
  year = {2018},
  doi = {10.1088/1475-7516/2018/05/035},
  eprint = {1803.01038},
  archivePrefix = {arXiv},
  primaryClass = {hep-ph},
  note = {A sparse numerical subset of the public EOS2018 table is supplied with this work}
}

\end{document}